\documentclass[fleqn,usenatbib]{mnras}

\usepackage{newtxtext,newtxmath}

\usepackage[T1]{fontenc}
\usepackage{ae,aecompl}

\usepackage{cheat_sheet}
\usepackage{url}
\usepackage{xspace}

\usepackage{graphicx}	
\usepackage{amsmath}	
\usepackage{amssymb}	
\usepackage{hyperref}   

\title[Analysis of SDSS spectra and GALEX photometry with {\sc starlight}]{Simultaneous analysis of SDSS spectra and GALEX photometry with {\sc starlight}: Method and early results}

\author[A. Werle et al.]{A. Werle,$^{1}$\thanks{E-mail: ariel@astro.ufsc.br}
 R. Cid Fernandes,$^{1}$
 N. Vale Asari,$^{1,2,3}$
 G. Bruzual,$^{4}$ 
 S. Charlot$^{5}$, \and
R. Gonzalez Delgado$^{6}$
and F. R. Herpich$^{1,7}$
\\
$^{1}$Departamento de F\'isica - CFM - Universidade Federal de Santa Catarina, Florian\'opolis, SC, Brazil\\
$^{2}$School of Physics and Astronomy, University of St Andrews, North Haugh, St Andrews KY16 9SS, UK\\
$^{3}$Royal Society--Newton Advanced Fellowship\\
$^{4}$Instituto de Radioastronom\'ia y Astrof\'isica, Universidad Nacional Aut\'onoma de M\'exico, Morelia, Michoac\'an, 58089 M\'exico\\
$^{5}$Sorbonne Universit\'es, UPMC-CNRS, UMR7095, Institut d'Astrophysique de Paris, F-75014, Paris, France\\
$^{6}$Instituto de Astrof\'isica de Andaluc\'ia (CSIC), Granada, Spain\\
$^{7}$Instituto de Astronomia, Geof\'{\i}sica e Ci\^{e}ncias Atmosf\'{e}ricas, Universidade de S\~{a}o Paulo, R. do Mat\~{a}o 1226, 05508-090 S\~{a}o Paulo\
, Brazil \\
}

\date{Accepted XXX. Received YYY; in original form ZZZ}

\pubyear{2015}

\begin{document}
\label{firstpage}
\pagerange{\pageref{firstpage}--\pageref{lastpage}}
\maketitle

\begin{abstract}
We combine data from the Sloan Digital Sky Survey and the Galaxy Evolution Explorer to simultaneously analyze optical spectra and ultraviolet photometry of 231643 galaxies with the \starlight\ spectral synthesis code using state-of-the-art stellar population models. 
We present a new method to estimate GALEX photometry in the SDSS spectroscopic aperture, which proves quite reliable if applied to large samples.
In agreement with previous experiments with CALIFA, we find that adding UV constraints leads to a moderate increase on the fraction of $\sim 10^7$ -- $10^8$ yr populations and a concomitant decrease of younger and older components, yielding slightly older luminosity weighted mean stellar ages.
These changes are most relevant in the low-mass end of the blue cloud. An increase in dust attenuation is observed for galaxies dominated by young stars.
We investigate the contribution of different stellar populations to the fraction of light in GALEX and SDSS bands across the UV-optical color-magnitude diagram.
As an example application, we use this $\lambda$ dependence to highlight differences between retired galaxies with and without emission lines. 
In agreement with an independent study by Herpich et al., we find that the former show an excess of intermediate age populations when compared to the later. 
Finally, we test the suitability of two different prescription for dust, finding that our dataset is best fitted using the attenuation law of starburst galaxies. However, results for the Milky Way extinction curve improve with decreasing $\tau_V$, especially for edge-on galaxies.

\end{abstract}


\section{Introduction}

The spectral energy distribution (SED) of galaxies encodes properties such as star-formation histories (SFH), stellar masses, metallicities, and dust attenuation. 
Stellar population synthesis techniques aim to extract this information by comparing the SEDs of galaxies with models. 

SED synthesis methods can be divided into two broad categories: parametric \citep{BEAGLE, BAGPIPES, CIGALE} and non parametric (or inverse) methods \citep{MOPED, Cid2005, VESPA, STECKMAP, ULySS, PPXF, Prospector}. 
The first compares galaxy spectra with a set of composite stellar population (CSP) models built by combining simple stellar population (SSPs) spectra according to prescribed star-formation and chemical enrichment histories.
In contrast, the latter retrieves stellar population information without any assumptions about the functional form of the galaxy's SFH  (see \citealt{Walcher2011} and \citealt{Conroy2013} for complete reviews).

The main advantage of parametric models is that they require less detailed spectral information to constrain physical properties. This allows them to be applied to broadband photometry, while the application of non-parametric population synthesis is usually restricted to $\lambda$-by-$\lambda$ spectral fits of the stellar continuum (i.e., excluding nebular emission).
For the same reason, certain galaxy properties such as abundance patterns and the contribution of binary stars to the SED can currently only be investigated through parametric models.
A particular asset of using broadband photometry to measure galaxy properties is that large photometric data-sets are publicly available across the entire electromagnetic spectrum, allowing for the analysis of panchromatic SEDs. Large databases of galaxy spectra, on the other hand, are only available in the optical region. 

To combine the best of both worlds, the \starlight\ spectral synthesis code \citep{Cid2005} was updated to simultaneously account for spectral and photometric information, allowing the analysis of panchromatic data while maintaining the detailed spectral constraints necessary for a non-parametric fit.
This method was first applied by \cite{Rafa2016} to simultaneously fit integrated spectra from the CALIFA survey \citep{CALIFA} and photometry from the Galaxy Evolution Explorer (GALEX, \citealt{Martin2007}).
We note that while \starlight\ is the first non-parametric code to include this feature, some parametric codes described in the literature also allow for the combination of spectra and photometry (e.g \textsc{beagle}, \citealt{BEAGLE}, and \textsc{bagpipes} \citealt{BAGPIPES}).

Several characteristics make the ultraviolet (UV) a natural choice for the expansion of the wavelength coverage of \starlight\ fits.
The main one is that this wavelength range provides key information about emission from OB stars, objects that dominate the UV emission in most galaxies even when they do not leave major footprints in the optical. Indeed, \cite{Rafa2016} find that \starlight\ fits to optical spectra alone tend to overestimate UV fluxes due to small ($\sim 2$\%) fractions of optical light attributed to these young, hot stars.  
The UV also opens a window to other astrophysical problems, such as the UV upturn in elliptical galaxies and the shape of dust attenuation curves.

This work aims to extend the analysis of UV constraints with \starlight\ to a combination of SDSS and GALEX data, taking advantage of a larger sample size. 
The paper is organized as follows: data sources and the procedure to match the data-sets are described in section \ref{sec:data}. Novelties in the spectral synthesis method are presented in Section \ref{sec:synthesis}. Synthesis results are discussed in Section \ref{sec:results}, with focus on the effect of UV data on the properties derived by \starlight. Examples of applications of the dataset to astrophysical problems are shown in Section \ref{sec:applications}. Finally, Section \ref{sec:conclusions} summarizes our results. Throughout this work, we assume a standard $\Lambda$CDM cosmology with $\Omega_{\rm M}=0.3$, $\Omega_\Lambda=0.7$ and $h=0.7$.

\section{Data and sample}\label{sec:data}

\subsection{Data sources and sample selection}

The work presented in this paper is based primarily on optical spectra from SDSS DR8 \citep{SDSS, DR8} and on UV photometry from GALEX GR6 \citep{Martin2007}, measured in two bands: $NUV$ (effective wavelength $\lambda^{\rm eff}_{\rm NUV} = 2267$ \AA) and $FUV$ ($\lambda^{\rm eff}_{\rm FUV} = 1516$ \AA). 

Our general sample contains 231643 galaxies from the SDSS main galaxy sample with photometry in both GALEX bands. The sample was selected from the GALEX CasJobs (\url{http://galex.stsci.edu/casjobs/}) by matching the UV object closest to each of the SDSS sources within a 0.3$^{\prime \prime}$ search radius. For most of the analysis, we use a subsample of 137979 galaxies with $z<0.1$, as this ensures that the wavelength ranges covered by the GALEX filters do not deviate largely from their rest-frame values; we will refer to this subsample as our low-$z$ sample.

SDSS spectra and the $NUV$ and $FUV$ magnitudes from GALEX constitute the main observational data for our analysis, but we also make use of SDSS $ugriz$ photometry to estimate GALEX magnitudes in the SDSS spectroscopic aperture, as explained next.

\subsection{Combining SDSS spectra and GALEX photometry}\label{sec:match}

The main limitation to our analysis is the FWHM of the GALEX PSF (4.6$^{\prime \prime}$ for $NUV$ and 5.4$^{\prime \prime}$ for FUV), which is larger than the aperture on which SDSS spectra are collected (1.5$^{\prime \prime}$ in radius). Our strategy to circumvent this issue is to indirectly estimate the GALEX 1.5$^{\prime \prime}$ magnitudes based on GALEX integrated photometry and the difference between total ($m_{\rm tot}$) and fiber ($m_{1.5}$) magnitudes $\Delta m_{1.5} \equiv m_{\rm tot} - m_{1.5}$ in the SDSS $ugriz$ bands.

The main factor that influences $\Dm$ is the fraction of the galaxy's area sampled by the fiber; a secondary but important factor is that $\Dm$ is larger in bluer filters, reflecting the difference in color between the central regions of galaxies (bulges) and their integrated light.   
We also find that $\Dm$ correlates weakly with galaxy morphology, and has no dependence on integrated color and absolute magnitude, indicating that $\Dm$ is not a strong function of galaxy type. Moreover, there is a tight correlation between $\Dm$ on different SDSS filters, indicating that the values of $\Dm$ in one band can be used to estimate $\Dm$ in other wavelengths.

\begin{figure}
 \centering
 \includegraphics[width=\columnwidth]{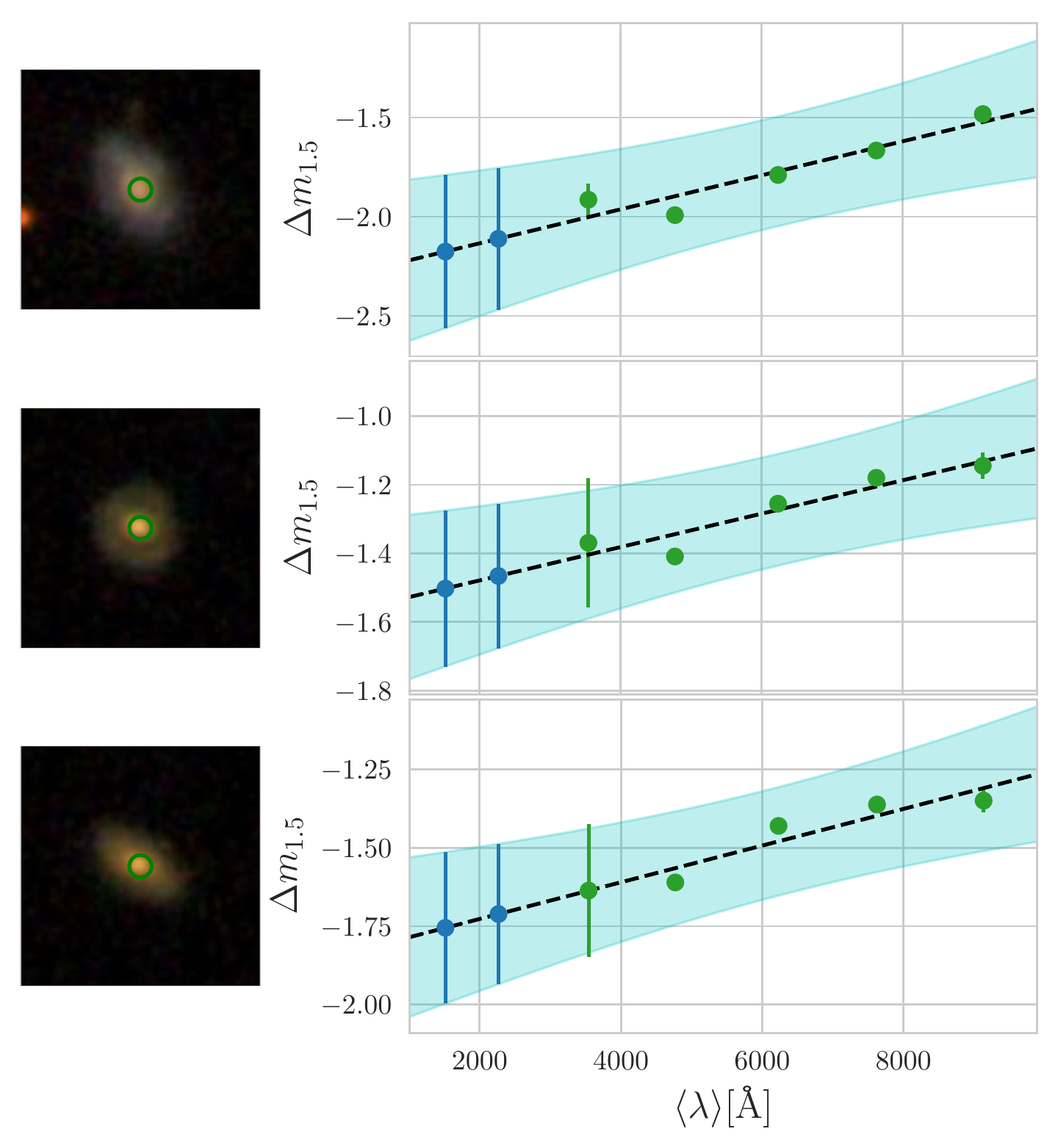}
 \caption{Examples of linear fits used to predict $\DmcalcNUV$ and $\DmcalcFUV$. Optical values of $\Dm$ are plotted in green  and the predicted values in blue. The best-fit linear relation is shown as a dashed line. The cyan band shows a $\pm 1 \sigma$ prediction band, used to estimate the uncertainties in $\DmcalcNUV$ and $\DmcalcFUV$. The estimated error bars are plotted in blue. An image of the corresponding galaxy is shown to the left of each fit, with the region covered by the SDSS fiber indicated by a green circle.}
 \label{fig:linear_example}
 \end{figure}

The relation of $\Dm$ with the mean wavelength of the filters $\langle \lambda \rangle$ is shown in Fig.\ \ref{fig:linear_example} for three example galaxies.
This relation was used to predict values for $\Dm$ in the ultraviolet by fitting a straight line to the $ugriz$ measurements and extrapolating it to the GALEX bands (dashed black lines). The lines were fitted using the BCES method \citep{BCES}, as implemented in python by \cite{Nemmen2012}. In this context, the offset  
of the fitted line acts as a first order correction that is equal for all bands and accounts for the fraction of total light sampled by the spectroscopic fiber, while the slope accounts for the aforementioned difference between central and integrated colors.

The UV magnitudes in the SDSS spectroscopic aperture are thus estimated from 

\begin{equation}\label{eq:aproxmags}
\begin{split}
 NUV_{1.5} = 
 NUV_{\rm tot} -  \widehat{\Delta} m_{1.5}^{NUV}  \\
 FUV_{1.5} =
 FUV_{\rm tot} -  \widehat{\Delta} m_{1.5}^{FUV} ,
\end{split}
\end{equation}

\noindent where 
$\widehat{\Delta} m_{1.5}^{NUV}$ and 
$\widehat{\Delta} m_{1.5}^{FUV}$ are the estimated corrections to the $NUV$ and $FUV$ bands. Notice that these corrections are estimated for each individual galaxy using its own $ugriz$ data. In these expressions $NUV_{\rm tot}$ and $FUV_{\rm tot}$ are the values of \texttt{NUV\_MAG\_AUTO} and \texttt{FUV\_MAG\_AUTO} in the GALEX catalog. The difference between kron photometry from GALEX and petrosian photometry from SDSS is very small and is therefore neglected.

Uncertainties in $\DmcalcNUV$ and $\DmcalcFUV$ are derived from the  $\pm 1\sigma$ prediction bands for the fitted lines, as indicated in Fig.\ \ref{fig:linear_example}. These aperture matching uncertainties are added in quadrature to the errors in $NUV_{\rm tot}$ and $FUV_{\rm tot}$ to obtain the error in $NUV_{1.5}$ and $FUV_{1.5}$.

\begin{figure}
 \centering
 \includegraphics[width=\columnwidth]{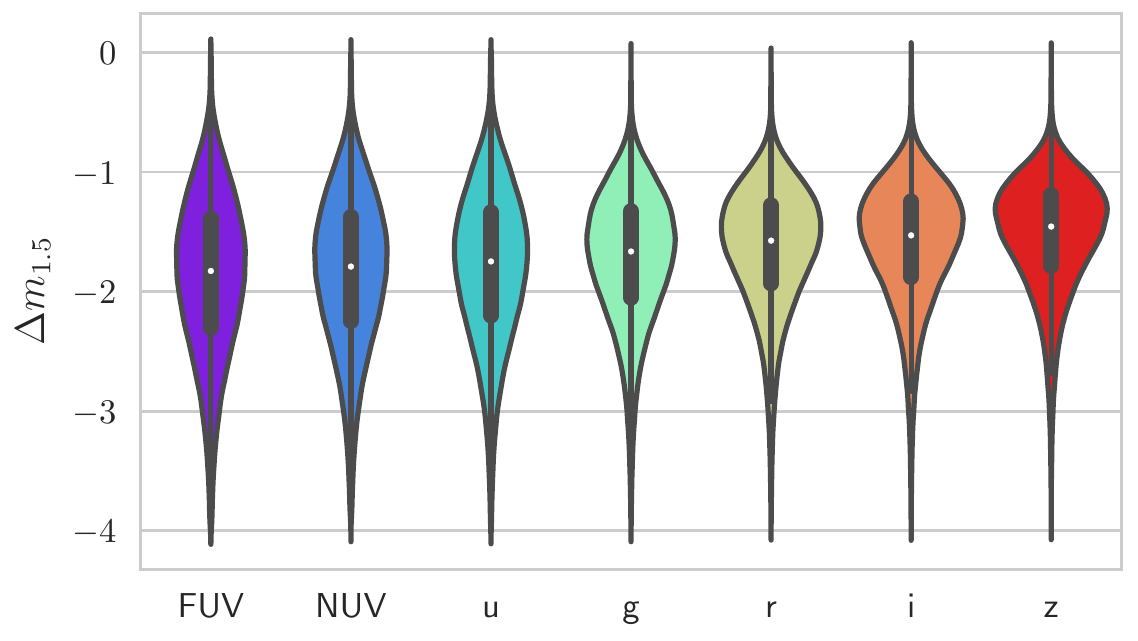}
 \caption{Violin plot of the distribution of $\Dm$ in different bands. The distributions show measured values for SDSS and calculated values for GALEX. Box plots marking the medians and quartiles are shown inside each distribution.}
 \label{fig:violin}
 \end{figure}

The distributions of $\DmcalcNUV$ and $\DmcalcFUV$ are compared to the observed $\Dm$ for the $ugriz$ bands in the violin plot in Fig \ref{fig:violin}. A box plot showing median and quartiles is plotted inside each distribution. The distribution of the calculated $\Dm$ for GALEX is very similar to the $u$-band,
and shifted to larger (more negative) values than for other bands.

The median aperture corrections for the sample are $-1.79$ mag for $NUV$ and $-1.82$ for $FUV$, while the median errors are 0.28 and 0.31 mag, respectively. 
We note, however, that since our method is completely unsupervised, it is expected to produce outliers, in particular for galaxies in close pairs or with odd morphological features. Indeed, in some cases the method returns unphysical values such as $\Dm>0$ (0.08\% of the sample), as well as very high error estimates (larger than 1 mag in the FUV band for 0.07\% of the sample).
Ultimately, the reliability of our method depends on the linearity of the $\Dm \times \langle \lambda \rangle$ relation. In cases where one or more bands deviate much from the linear trend the method is not reliable and will return high error estimates.

\begin{figure}
 \centering
 \includegraphics[width=\columnwidth]{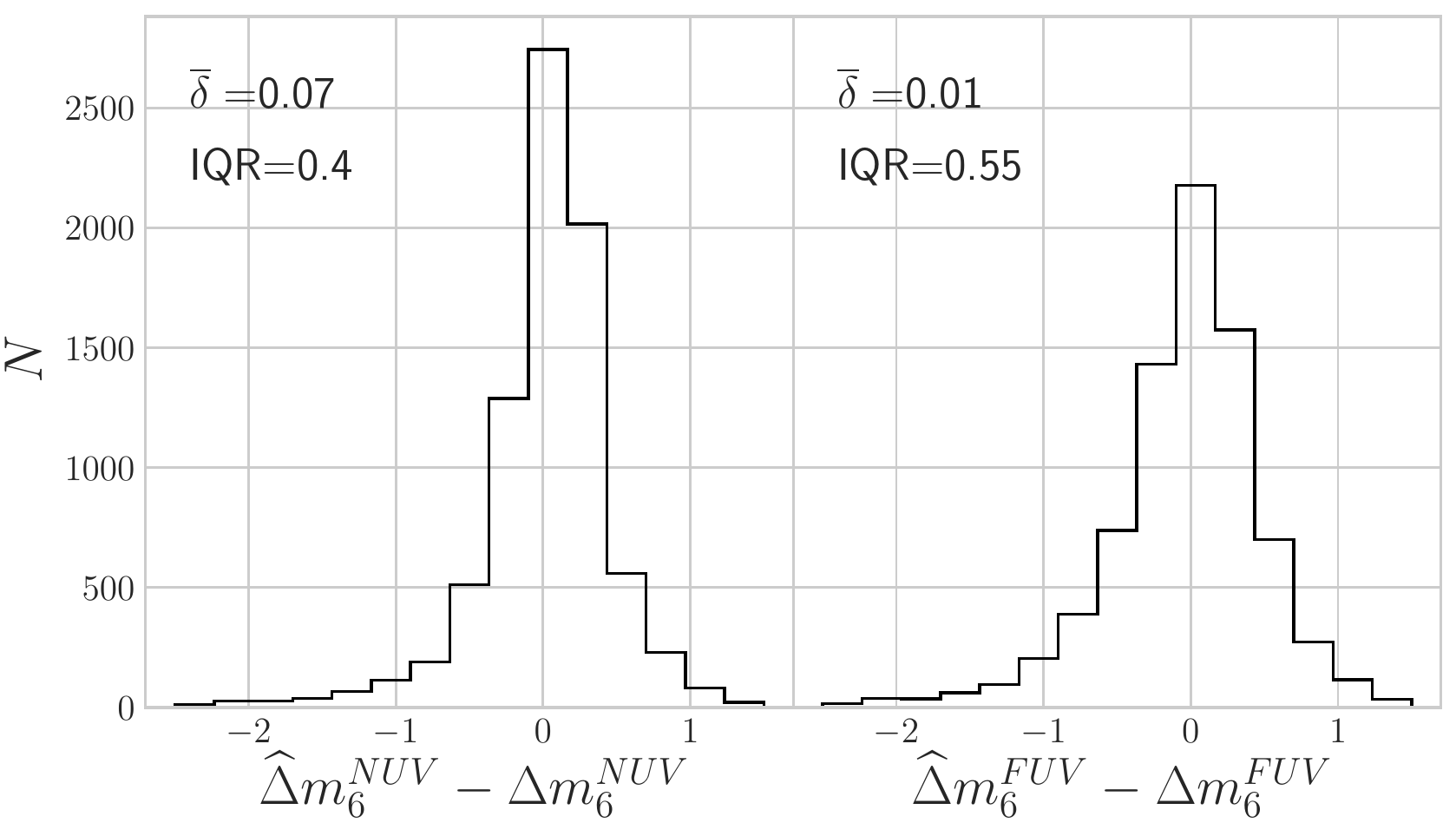}
 \caption{Comparison between predicted and measured values of $\DmNUVs$ and $\DmFUVs$ for a test sample of 6105 galaxies from the NASA-Sloan Atlas.
Histograms show the differences between estimated and observed $\Dms$ for $NUV$ (left) and $FUV$ (right) bands. Medians ($\overline{\delta}$) and interquartile regions (IQR) of the distributions are annotated in each panel.}
 \label{fig:test}
 \end{figure}

In order to test the reliability of our method, we use the GALEX-SDSS astrometry-matched radial profiles available in the NASA-Sloan Atlas \citep[NSA,][]{Blanton2011} to calculate the differences between the total magnitudes and that in an aperture of 6$^{\prime\prime}$ in radius ($\Dms$), where the image degradation caused by the GALEX PSF can be neglected.  
The idea behind this test is that $\Dm$ can be thought of as the log of the ratio between total flux and the flux in the fiber, which is a relative quantity. Therefore, by choosing a larger aperture we can use nearby galaxies (with large projected radii) to mimic the distribution of $\Dm$ in the general sample.
With that in mind, we selected a test sample of 6105 galaxies with values of $\Dms$ in the SDSS bands in the same range as the $\Dm$ values for the general sample.
Our method was then applied to estimate the values of $\Dms$ for the test sample.

Results of this test are shown in the histograms of Fig. \ref{fig:test}, where we plot the distributions of $\DmcalcNUVs-\DmNUVs$ and $\DmcalcFUVs-\DmFUVs$. The distributions are centered near zero, with medians of 0.07 mag for $NUV$ and $-0.01$ for $FUV$. However, both distributions are quite wide, with interquartile regions of $0.4$ on $NUV$ and $0.55$ on $FUV$. These statistics confirm that our estimates are good for large samples, although results for individual sources may not be reliable.  

\begin{figure}
 \centering
 \includegraphics[width=\columnwidth]{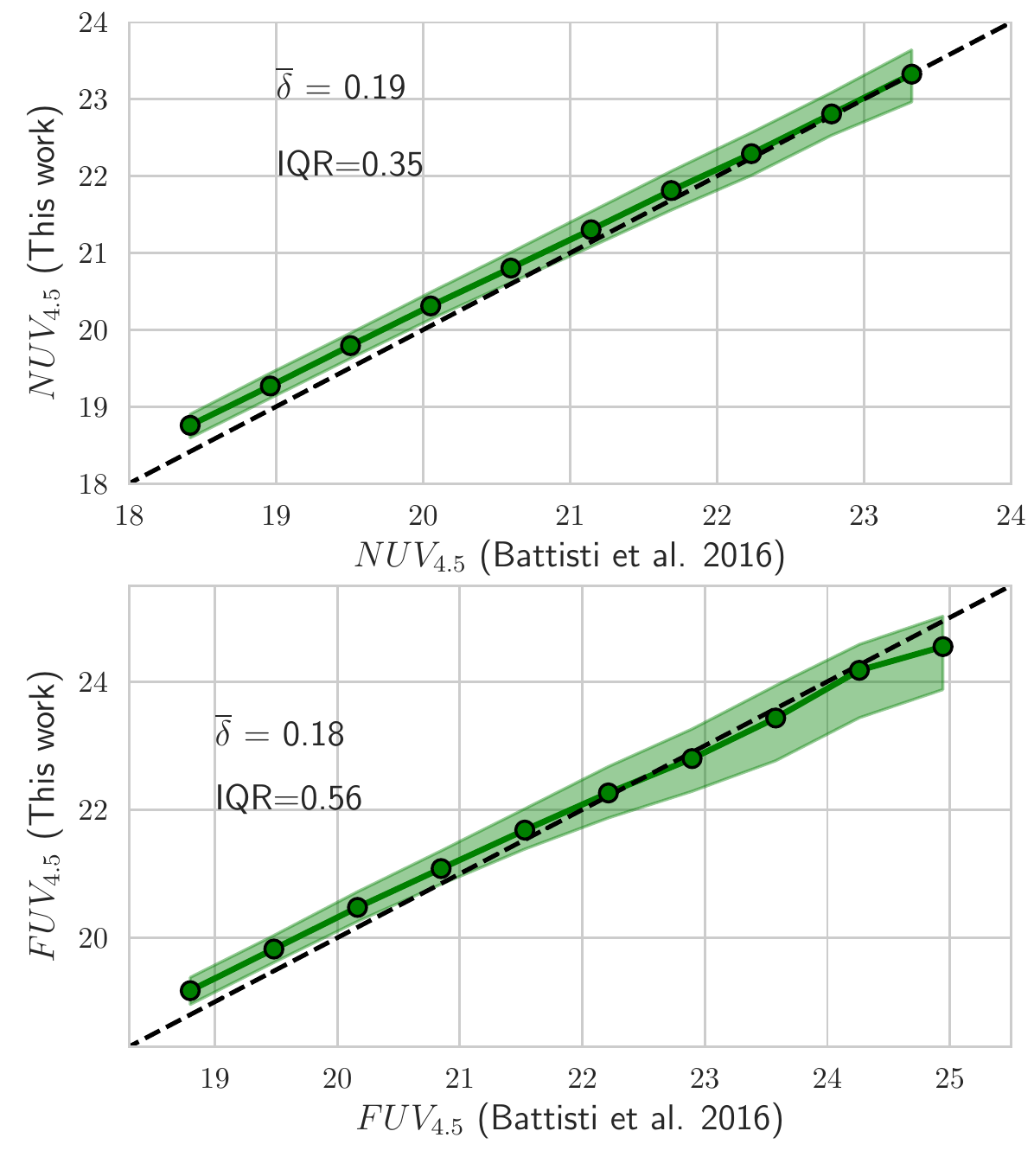}
 \caption{Comparison of UV magnitudes calculated by \protect\cite{Battisti2016} with UV magnitudes calculated using the method of this paper for an aperture of radius 2.25''. Green lines show median curves with the region between 25 and 75\% percentiles is shaded in green. Median $y-x$ ($\overline{\delta}$) values and interquartile regions (IQR) are annotated on each panel.
 }
 \label{fig:test_BCC}
 \end{figure}

We also used the NSA to calculate $\Delta m$ in an aperture of  2.25$^{\prime \prime}$ in radius, 
which allows us to compare our aperture matching scheme to the one of \cite{Battisti2016}, that is calibrated in this aperture. This comparison is shown in Fig. \ref{fig:test_BCC}. Magnitudes obtained with both methods are linearly correlated, although the method described here produces systematically dimmer magnitudes.
Over the whole sample, the median difference between the methods is 0.19 magnitudes on $NUV$ and 0.18 on $FUV$, the interquartile regions (IQR) are of 0.35 magnitudes on $NUV$ and 0.56 on $FUV$, with a larger scatter for dimmer sources, specially in the $FUV$ band.
Considering that these two aperture matching schemes follow completely different routines and are even based on different data\footnote{The correction from \cite{Battisti2016} uses a convolution of the galaxy's S\'ersic profile in the $u$ band with the GALEX PSF to calculate scale factors that are applied to the GALEX 1.5$^{\prime\prime}$ photometry.}, the agreement between the predicted magnitudes is reassuring for both methods.

Despite the wide distributions of $\widehat{\Delta} m_6^{NUV}-\Delta m_6^{NUV}$ and $\widehat{\Delta} m_6^{FUV}-\Delta m_6^{FUV}$, the predicted values are, on average, in very good agreement with the observed ones. This is sufficient for the purpose of this paper, since all our results will be averaged in relatively large samples. Nonetheless, in some cases this process can yield corrections that deviate from the distributions shown in Fig \ref{fig:violin}. To circumvent this we remove galaxies with corrections larger than 0 or smaller than -4 magnitudes in one of the GALEX bands, discarding 3300 galaxies (0.14\% of the sample).

\subsection{Preprocessing steps}

Besides the aperture matching correction explained above, the data undergo a few pre-processing steps before being fed to {\sc starlight}.

From DR7 onwards, SDSS spectra are calibrated to match the flux of a point source within one FWHM of the PSF, leading to  fluxes typically 25\% smaller than their photometric counterparts for extended sources. Due to this offset, calculating $NUV$ and $FUV$ fluxes compatible with fiber photometry does not directly ensure compatibility with spectroscopic  fluxes. We have thus rescaled the SDSS spectra to match the 3'' photometry in the $r$ band, as done in the MPA/JHU value added galaxy catalog (\url{wwwmpa.mpa-garching.mpg.de/SDSS/DR7}). Since the shape of the photometric SEDs is very similar to the shape of the spectra, this procedure ensures the compatibility of SDSS spectra with $ugriz$ fiber photometry and, by extension, with the estimated $NUV_{1.5}$ and $FUV_{1.5}$ estimated fiber magnitudes. 

Both the optical spectra and UV photometry are corrected for Galactic extinction using the \cite*{CCM} extinction law with $R_V = A_V/E(B-V) = 3.1$, and $E(B-V)$ values from \cite{SFD}  assuming the recalibration introduced in \cite{Schlafly2011}. 

Finally, the spectra are shifted to $z = 0$ and re-sampled to 1\AA\ wavelength intervals. UV fluxes, however, are not k-corrected to rest-frame values. Instead, \starlight\ uses the known redshift of the source to evaluate the predicted $NUV$ and $FUV$ magnitudes in the observed frame.

\subsection{Ancillary data}

The stellar population analysis presented in this work is based on the pre-processed SDSS spectra and GALEX magnitudes described in the previous sections. Even though no other data is required for this analysis, information on properties like emission lines and host morphology is key to the interpretation of the results. Following the statistical gist of this study, we will use such ancillary data to organize sources into different groups and examine the results in a comparative way.
We use the Galaxy Zoo \citep{GZ} morphology to define subsamples of ellipticals and spirals. Emission line fluxes used in section \ref{sec:Lynis_x_Lineless} are drawn from \cite{Abilio2006}.

The main method to group galaxies explored in this paper is the $NUV - r$ versus $M_r$ color magnitude diagram (CMD). Unlike purely optical based CMDs, where the red sequence, blue cloud and green valley populations exhibit substantial overlap, this UV-optical CMD clearly represents the bimodality of the local galaxy population, making it a valuable tool to examine trends for different galaxy types \citep{Martin2007, Salim2014, Thiago2012}. 
To build this CMD we calculated absolute magnitudes, which are corrected for Galactic extinction and k-corrected to $z=0$  using the \textsc{kcorrect} code \citep{kcorrect}. We note that this is the only use of k-correction in this paper.

\section{Spectral Synthesis}\label{sec:synthesis}

\subsection{Method}

The version of \starlight\ used in this paper introduces a few novelties with respect to \cite{Rafa2016}, the main change being the scheme for combining photometric and spectroscopic figures of merit. 

%
%

The spectroscopic and photometric parts of the $\chi^2$ are defined as 

\begin{equation}
\begin{split}
\chi^2_{\rm SPEC} & = \sum\limits_{\lambda}  w_\lambda^2 (O_\lambda - M_\lambda )^2 \\
\chi^2_{\rm PHO} & = \sum\limits_{l=0}^{N_l} \left( \frac{m_l^{\rm obs} - m_l^{\rm mod}}{\sigma_l}\right)^2,
\end{split}
\end{equation}

\noindent where $O_\lambda$ and $M_\lambda$ are the observed and model spectra, $w_\lambda$ is the inverse error in $O_\lambda$ (except in bad pixels and regions around emission lines, which are discarded from the fits by setting $w_\lambda = 0$), $l$ is an index corresponding to each of the $N_l$ photometric filters, $m_l^{\rm obs}$ and $m_l^{\rm mod}$ are the observed and modeled AB magnitudes in filter $l$ and $\sigma_l$ are the errors in the magnitudes. 
Model magnitudes are calculated by performing synthetic photometry in the model spectrum $M_\lambda$ after shifting it to the observed frame, ensuring consistency with the input apparent magnitudes $m_l$.

In all practical applications, $N_l$ is much smaller than the number of spectroscopic data points $N_\lambda$ -- for instance, $N_l = 2$ and $N_\lambda \sim 4000$ in this paper.  In order to simultaneously minimize $\chi^2_{\rm SPEC}$ and $\chi^2_{\rm PHO}$, this scale difference has to be circumvented in our definition of the total $\chi^2$. Therefore, we define the total $\chi^2$ as

\begin{equation}
 \chi^2_{\rm TOT} = \chi^2_{\rm SPEC} + g_{\rm PHO} \frac{ N_\lambda }{ N_l }\chi^2_{\rm PHO},  
\end{equation}

\noindent where $g_{\rm PHO}$ is a technical parameter that sets the relative weight of photometric and spectroscopic constraints.

\cite{Rafa2016} did experiment with this recipe, but concluded that no scaling of $\chi^2_{\rm PHO}$
was necessary to adequately fit both their UV photometry and optical spectra. This is not, however, valid in general. In particular, the uncertainties in the spectroscopic and photometric fluxes play a central role in defining the balance between $\chi^2_{\rm SPEC}$ and $\chi^2_{\rm PHO}$. Optimal values of $g_{\rm PHO}$ are thus dataset-dependent.

For the combination of SDSS spectra and aperture-corrected GALEX photometry used in this paper we find that $g_{\rm PHO}$ values of the order of 0.1 are necessary to adequately fit both the UV and optical data, so this is the value adopted throughout this work. Increasing $g_{\rm PHO}$ improves the fit of UV fluxes at the expense of degrading the quality of the fit of the optical spectrum (particularly in its blue end), and vice versa.

\subsection{Stellar population base}

Spectral synthesis with \starlight\ depends on a few astrophysical ingredients, the main one being a base of stellar population spectra.

The base used in this paper is built from SSP models from Charlot \& Bruzual (2018, in preparation, CB18 hereafter). 
which include considerable improvements upon the \cite{BC03} models.
The CB18 models incorporate the PARSEC evolutionary tracks computed by \cite{Chen15} for 16 values of the stellar metallicity ranging from $Z = 0$ to $Z = 0.06$. 
These tracks include the evolution of the most massive stars losing their hydrogen envelope through the Wolf-Rayet (WR) phase, and have been complemented with the work by \cite{Marigo13} to follow the evolution of stars through the thermally pulsing asymptotic giant branch (TP-AGB) phase.
A large number of empirical and theoretical stellar libraries is used to describe the spectrophotometric 
properties of the stars along these tracks. 
For the age and wavelength ranges of interest for this paper, the dominant stellar spectra come in the
visible range from the MILES stellar library \citep{sanchez2006,falcon2011} and in the UV range from the
theoretical libraries computed by 
\cite{lanz2003,lanzerr2003,lanz2007,leitherer2010,martins2005,rodmer2005,rauch2003}, and the high
resolution PoWR models
\citep{sander2012,hamman2006,hainich2014,hainich2015,todt2015,graefener2002,hamman2003} to describe
stars in the WR phase.
The effects of dust shells surrounding TP-AGB stars on their spectral energy distribution
\citep{aringer2009,rayner2009,westera2002} is treated as in \citet{RAGL10}.
The CB18 models have been used, among others, by \cite{Gutkin16,Wofford16,Vidal17,fritz2017ApJ,bitsakis2017,
bitsakis2018}.
These models are available to the interested user upon request.
A comparison between the new models and the ones from \cite{BC03} is made in appendix A of \cite{Vidal17}. A more detailed comparison will be provided by CB18.

We use these SSP models to compute the spectra of CSPs resulting from periods of constant star formation rate. The age range from $t = 1$ Myr to 14 Gyr was divided onto 16 logarithmically spaced age bins. We use seven values of metallicity between $Z = 0.0005$ and $3.5 Z_\odot$  ($Z_\odot = 0.017$), yielding a base of $16 \times 7 = 112$ CSPs.

\subsection{Dust attenuation}

The effects of dust attenuation
\footnote{Throughout this paper, we adopt the standard definition of "extinction" as the scattering and absorption of photons out of the line of sight, while "attenuation" is defined as the combination of absorption and scattering in and out of the line of sight with local and global geometric effects. } in this work are modeled as if produced by a single foreground screen which attenuates fluxes by a factor $e^{-\tau_\lambda}$, with $\tau_\lambda$ parameterized as the product of the optical depth in the V-band ($\tau_V$) and a $q_\lambda \equiv \tau_\lambda / \tau_V$ attenuation or extinction curve.

We have performed spectral fits with two different recipes for $q_\lambda$:  the MW law, as parameterized by \cite{CCM} with $R_V=3.1$ (CCM), and the \cite{Calzetti2000} law (CAL), modified in the $\lambda < 1846$ \AA\ range to smoothly turn into the curve derived by \cite{Leitherer2002}.\footnote{This modification has only minor effects, restricted to the $FUV$ band.} 
The most notable difference between these laws is the presence or absence of the so called `UV bump', a broad peak in $q_\lambda$ around 2175 \AA\ \citep{Stecher1965}, within the range of the $NUV$ filter.
A more fundamental difference between them is that while CAL formally represents an attenuation law, CCM is originally an extinction curve, a crucial distinction when modeling the two processes (e.g \citealt{Witt2000}).
In the context of this work, however, CCM is used to model attenuation, so the term `attenuation law' may loosely apply.
These two laws are chosen because they are the most often used in \starlight-based work, and also because they allow us to test how an UV bump affects the fitting of UV data. 
These characteristics suit our central goal, which is to showcase the potential of the combined analysis of optical spectra and UV photometry with \starlight.

A comparison of results obtained with the two laws is presented in section \ref{sec:dustlaws}, but we anticipate that the best results for the general population of galaxies are obtained for the Calzetti law. Accordingly, all results presented up to section \ref{sec:dustlaws} assume this prescription.

\section{Synthesis Results}\label{sec:results}

Having discussed the sample, how we handle the data, and the method of analysis, this section presents the results of the synthesis. Throughout this section the emphasis is on the comparison of results obtained from purely optical spectral fits to SDSS data with those obtained with the addition of GALEX $NUV$ and $FUV$ photometry.

\subsection{Spectral fits and UV magnitudes}\label{sec:output}

\begin{figure}
 \centering
 \includegraphics[width=\columnwidth]{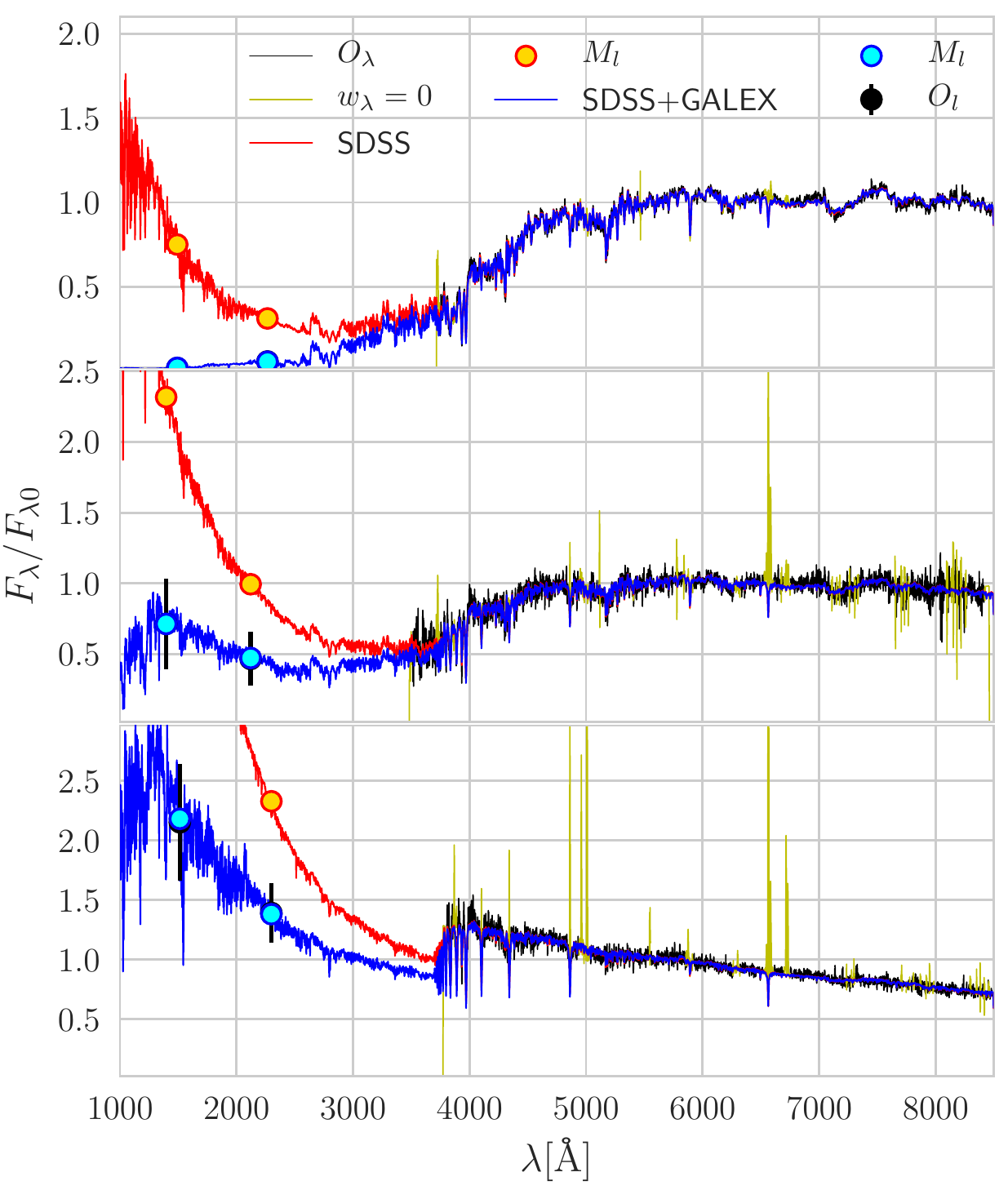}
 \caption{Examples of \starlight\ fits for three galaxies in our sample. Black lines show optical spectra, with masked or flagged regions marked in green. Red lines show purely spectroscopic fits, while blue lines show combined  UV + optical fits.
 Black circles with error bars mark the observed GALEX fluxes ($NUV_{1.5}$ and $FUV_{1.5}$, scaled to the SDSS aperture using the method described in Section \ref{sec:match}), cyan circles are the fitted GALEX fluxes. Orange circles show UV fluxes predicted from purely optical fits. All fluxes are relative to the flux at 5635 \AA.}
 \label{fig:example_fits}
 \end{figure}

Fig. \ref{fig:example_fits} show examples of spectra fitted with and without photometric constraints for three different galaxies, ordered from red (top) to blue (bottom) $NUV-r$ colors.
The observed SDSS spectra are shown in black, except for regions masked because of bad pixels or emission lines, which are shown in yellow. Red and blue lines show the optical and combined optical + UV  model fits, respectively.  Observed UV fluxes are shown as black circles, while model values are plotted as orange circles for the optical-only fits and cyan for the optical + UV fits. All fluxes are relative to the flux at the normalization wavelength, set to $\lambda_0 = 5635$\AA.

Fig. \ref{fig:example_fits} confirms the results of \cite{Rafa2016}, in that purely optical fits tend to predict UV fluxes much larger than the observed ones. 
This happens because, as will be discussed next, optical-only fits can easily accommodate small contributions of very young stellar populations which hardly affect the optical fluxes, but become dominant at UV wavelengths. 
The addition of UV constraints allows \starlight\ to successfully fit UV magnitudes with very little changes to the fitted optical spectra.
Also, since this effect involves only small fractions of stellar mass and optical light, no drastic change is observed in the measured star-formation histories, as will become clear in Figs.\ \ref{fig:SFH} and  \ref{fig:SFHNUVr}. 

 \begin{figure}
 \centering
 \includegraphics[width=\columnwidth]{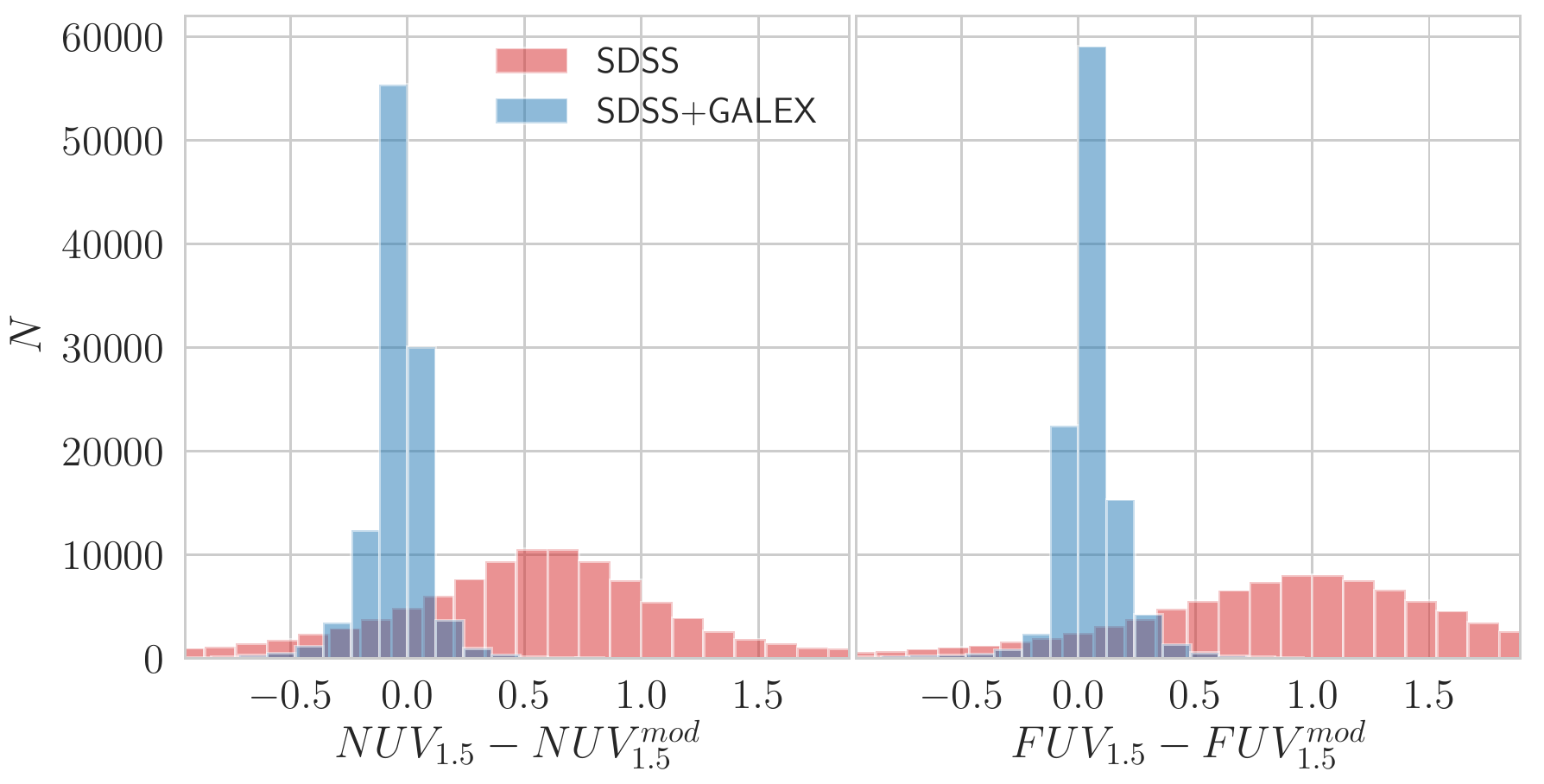}
 \caption{Histograms comparing differences between observed $NUV$ (left) and $FUV$ (right) magnitudes and the values predicted from optical spectra (red) and fitted from GALEX photometry (blue).}
 \label{fig:hists}
 \end{figure}

This effect is further illustrated in  Fig \ref{fig:hists}, where we plot histograms of the difference between the modeled and observed UV magnitudes for the full sample.
For fits without photometric constraints, UV magnitudes are overshot by $0.41\pm0.91$ magnitudes on $NUV$ and $0.83\pm1.32$ on $FUV$, yielding values that are often brighter than the integrated UV magnitudes.

\subsection{Star Formation Histories}

We now examine the changes in the derived SFHs generated by the inclusion of UV constraints in the analysis. Within \starlight\ SFHs are described in terms of the light fraction population vector $\vec{x}$, which quantifies the contribution of different base components to the observed flux at a normalization wavelength. The mass fraction vector $\vec{\mu}$ is derived from $\vec{x}$ and the light-to-mass ratios of the base populations.

\begin{figure}
 \centering
 \includegraphics[width=\columnwidth]{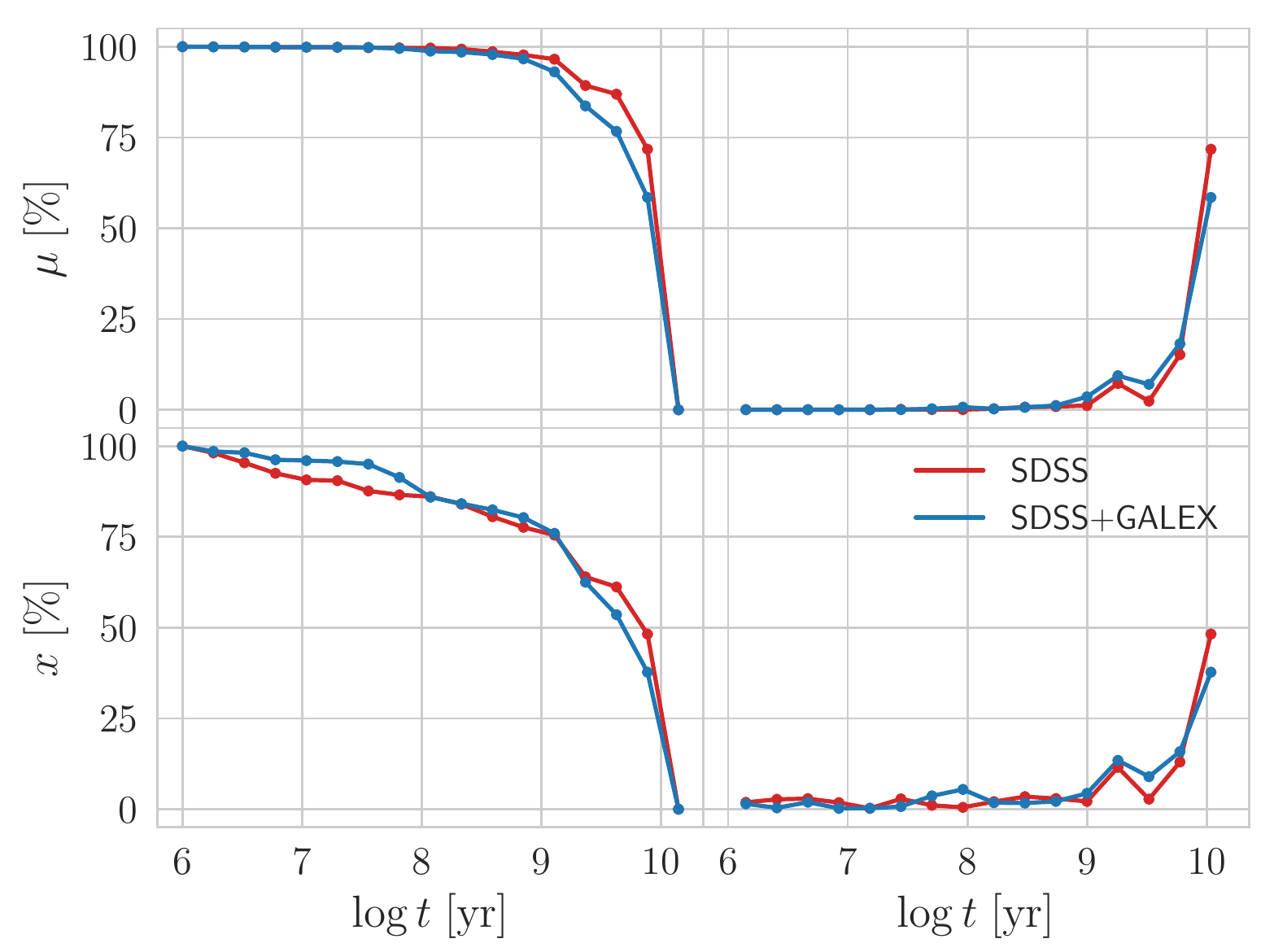}
 \caption{Average star formation histories calculated with and without photometric constraints. Top panels show cumulative (left) and non-cumulative (right) mass fractions as a function of age, while bottom panels show the corresponding curves in terms of light fractions at $\lambda_0 = 5635$\AA. Red lines show values for SDSS only \starlight-fits and blue lines show values for SDSS+GALEX.
 }
 \label{fig:SFH}
 \end{figure}

Fig.\ \ref{fig:SFH} shows the average $\vec{x}$ and $\vec{\mu}$ vectors for all galaxies in our low $z$ sample, obtained after collapsing the $Z$-axis (i.e., adding all components with same age but different metallicities). 
These distributions of mass and light with age are shown in both fractional (right panels) and cumulative  forms (left).

The addition of UV data shifts light fractions from the $<10^7$ yr populations to slightly older ones, from $10^7$ to $10^8$ yr, as can be seen in the bottom panels of Fig. \ref{fig:SFH}. This is enough to prevent the overshooting of UV fluxes, but tends to produce a redder optical spectrum. To prevent this, \starlight\ also removes contributions from the oldest stellar populations.
The $\chi^2_{\rm SPEC}$ values obtained with the addition of UV constraints are only marginally worse (by 3\% on average) than those obtained with purely optical fits, confirming that the fitted optical spectra are kept essentially unchanged.
When translating light fractions into mass fractions, the main difference between the two types of fits lies in the older populations: Fits with UV constraints show a slower and smoother build-up of stellar mass at early epochs.

\begin{figure*}
 \centering
 \includegraphics[width=\textwidth]{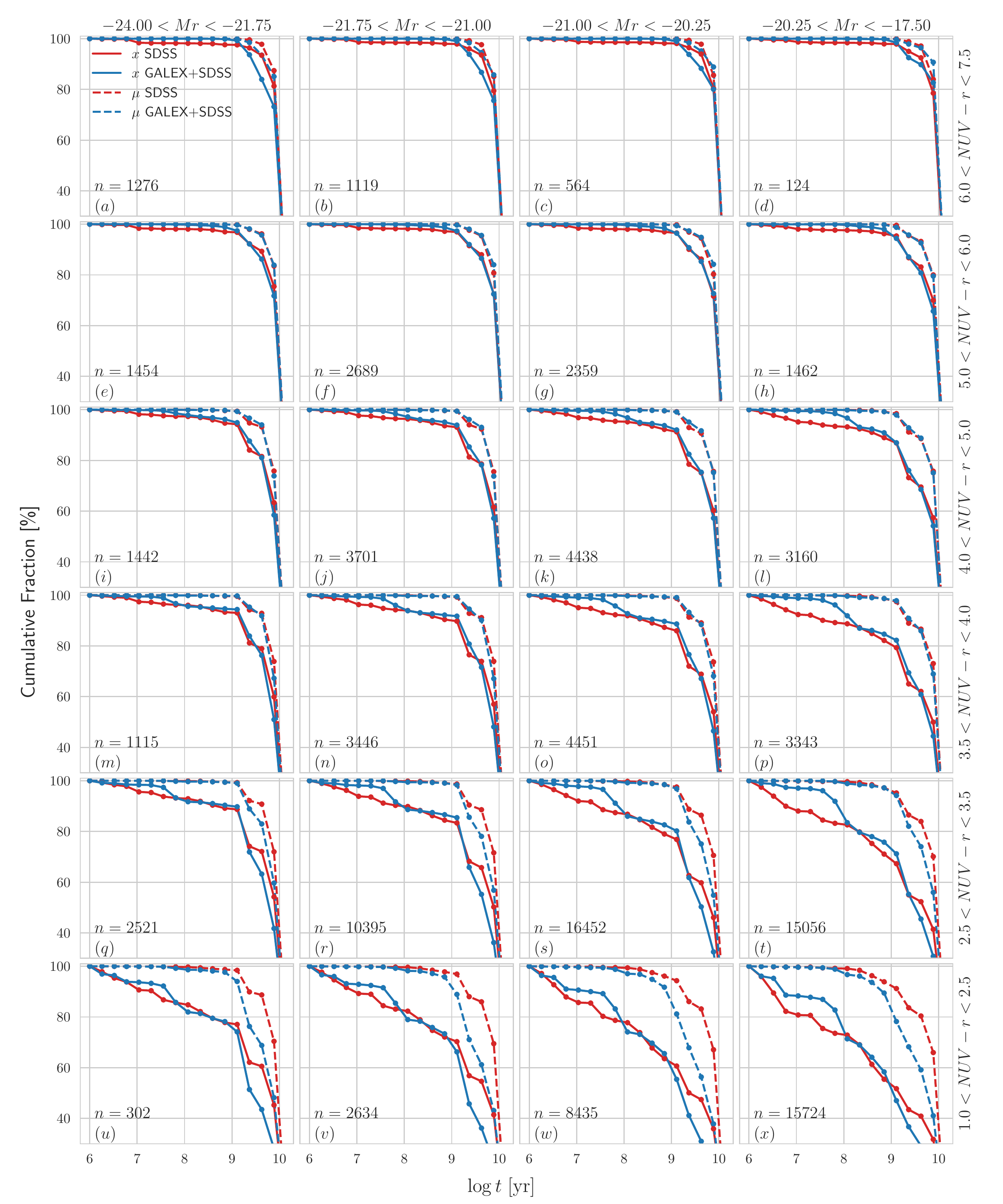}
 \caption{Average star-formation histories calculated with and without photometric UV constraints across the $NUV - r$ vs.\ $M_r$ color-magnitude diagram. Solid lines show cumulative light fractions at $\lambda=5635$\AA\ while dashed lines show cumulative mass fractions. Red lines show results for pure SDSS fits and blue lines show results for combined GALEX+SDSS fits. Dots mark the edges of the CSP age bins. The number of galaxies in each bin ($n$) is indicated in the subplots.
 }
 \label{fig:SFHNUVr}
 \end{figure*}

The sample averaged SFHs in Fig.\ \ref{fig:SFH} mix very different kinds of galaxies, however. To get a sense of the diversity of SFHs in our sample and how they change with the addition of UV data, we calculated average light and mass fractions in bins across the $NUV - r$ vs. $M_r$ color-magnitude diagram. In this diagram, and at the redshift limit of our sample, the blue cloud can be defined by the criterion of $NUV-r<4$ and the red sequence by $NUV-r>5$, while points in the $4<NUV-r<5$ range lie in the green valley \citep{Salim2014}. On the assumption that sources in a given locus of the CMD can be considered intrinsically similar, these local averages are much less sensitive to sample selection effects. The results are plotted in Fig.\ \ref{fig:SFHNUVr}.

The largest changes in SFH occur for galaxies in the low-mass end of the blue cloud (bottom-right panels), where young stellar populations are more abundant and UV constraints are expected to play a larger role. The redistribution of $< 10^7$ and $> 10^9$ to $10^7$--$10^8$ yr populations, which was the most noticeable change for the sample average, is very clear among these galaxies.

We also note that purely optical fits of red sequence galaxies (top panels) tend to wrongly identify very young populations ($t<10$ Myr) at levels of order 2\%. These are the fake young bursts first identified by \cite[see also \citealt{Cid2010}]{Ocvirk2010}. The addition of UV data constrains these populations, thus averting the problem.

Overall, the differences between SFHs calculated with and without UV data become smaller with redder colors and increasing luminosity, i.e., as the contribution of old stars increase.
For light fractions, our results are similar to the ones obtained by \cite{Rafa2016}. For mass fractions, however, there are differences. While \cite{Rafa2016} finds faster rising cumulative $\mu(>t)$ curves for optical+UV fits than with purely optical ones in blue cloud galaxies, here we observe a trend in the opposite direction (as best seen in panel w of Fig.\ \ref{fig:SFHNUVr}). Besides the many differences in sample, data, aperture corrections, and optical/UV weighing scheme, the current fits differ in the base models, which employ different isochrones and stellar libraries. We have verified that using base models compatible with those used by \cite{Rafa2016} moves our $\mu(>t)$ curves closer to theirs. Limiting the optical fits to the same 3800--7000 \AA\ range of their spectra further improves the agreement.

\subsection{Global properties}

\begin{figure*}
 \centering
\includegraphics[width=\textwidth]{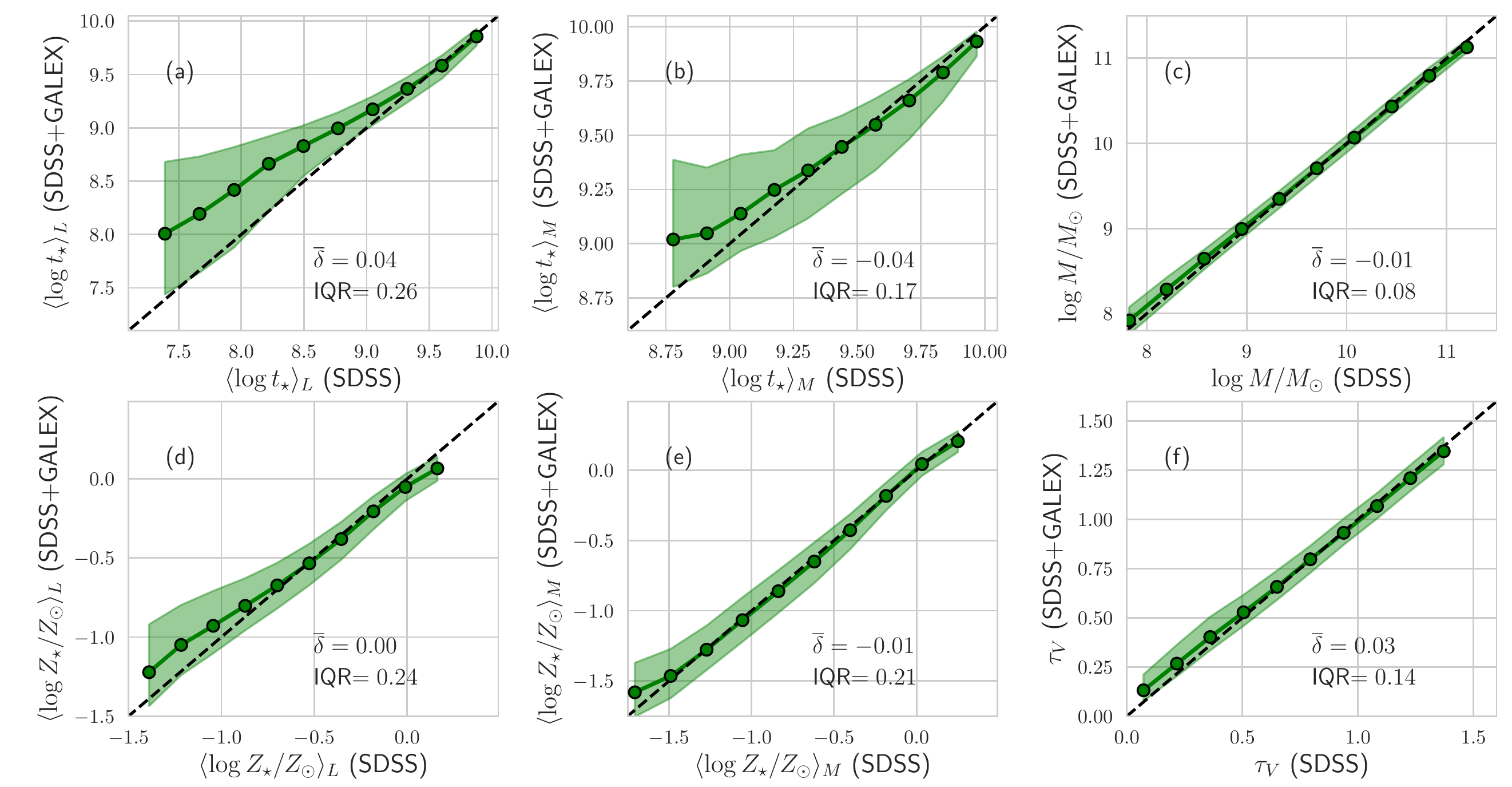}
 \caption{Comparison between galaxy properties derived with and without UV data. The $y$-axis corresponds to fits with UV constraints and the $x$-axis to fits to optical spectra, green lines show median curves with the center of each bin marked as a green dot with a black border. The region between 25 and 75\% percentiles is shaded in green. Panels show mean stellar ages weighted by light (a) and mass (b), stellar masses (c), light (d) and mass (e) weighted mean stellar metallicities and V-band dust optical depth (f). Dashed black lines show a $y=x$ relation. Median $y-x$ values ($\overline{\delta}$) and interquartile (IQR) regions are annotated on each panel.
 }

\label{fig:OPTxPHO}
\end{figure*}

In addition to the population vectors discussed above, \starlight\ also returns global properties such as mass and dust attenuation; other properties like mean stellar ages and metallicites can be calculated by reducing the dimensionality of the population vector.
Fig.\ \ref{fig:OPTxPHO} compares a series of these global properties derived with ($y$-axis) and without ($x$-axis) UV photometry. In order to highlight the general trends throughout the sample and give less weight to outliers, results are plotted as median curves with the region between 25 and 75\% percentiles shaded in green.
$\overline{\delta}$ and IQR values in each panel denote the median (bias) and interquartile region (scatter) of the difference between $y$ and $x$ values.

Panels (a) and (b) compare luminosity (at $\lambda=5635$\AA) and mass weighted mean stellar ages, 
$ \langle \log t \rangle_L$ and $\langle \log t \rangle_M$, respectively.
In the case of $\langle \log t \rangle_L$ the bias is slightly positive  ($\overline{\delta} = +0.04$ dex), and driven by the youngest systems, i.e., blue cloud galaxies, whose youngest populations shift from the $\log t = 6$--7 range to the next decade ($\log t = 7$--8) when UV constraints are used in the fits (see Fig.\ \ref{fig:SFHNUVr}).
Mass weighted log ages (Fig.\ \ref{fig:OPTxPHO}b) are less sensitive to the recent SFH, thus spanning a much smaller range. Still, the negative bias of $\overline{\delta} = -0.04$ dex in $\langle \log t \rangle_M$ reflects how UV data leads \starlight\ to bring down the contribution of the oldest stellar populations.

Panels (d) and (e) in Fig.\ \ref{fig:OPTxPHO} show luminosity and mass weighted log metallicites, respectively. UV data makes the most metal-poor galaxies become slightly less so, although overall biases is zero for $\langle \log Z \rangle_L$ and negligible for $\langle \log Z \rangle_M$ ($\overline{\delta} = -0.01$ dex). \cite{Rafa2016} find a larger negative bias in $\langle \log Z \rangle_M$, mostly due to late type, blue cloud galaxies, whose metallicities come out smaller in UV+optical fits. Again, this difference is due to differences in both data (mainly spectral coverage in the optical)  and ingredients in the analysis (base models). Repeating the analysis mimicking their setup we reproduce their results. 

UV constraints do not affect the estimates of stellar masses, as seen in Fig.\ \ref{fig:OPTxPHO}(c). This is expected, given that we have seen that UV data mostly affects the youngest populations, which carry little mass. 
Somewhat more surprisingly, but in agreement with  \cite{Rafa2016},  Fig.\ \ref{fig:OPTxPHO}f shows that, in general, optical and UV+optical fits produce similar estimates of the dust attenuation, here converted to dust optical depth $\tau_V$. This counter-intuitive result will be dissected in the following section.

The median curves in Fig. \ref{fig:OPTxPHO} highlight the general trends throughout the sample. As expected, there are points that fall out of this trend. This is the case for some galaxies with blue $NUV-r$ and relatively red $FUV-NUV$, a combination of colors that is sometimes achieved by reddening very young populations. When this effect is in place, the addition of UV data makes galaxies significantly younger and more reddened, deviating from the relations shown in panels (a), (b) and (f) of Fig. \ref{fig:OPTxPHO}. This only happens for a small population of galaxies that bears no effect to the general trend, although this behavior is an interesting clue on galaxies that require two components of dust attenuation, which in principle could yield redder $FUV-NUV$, while having a smaller effect in $NUV-r$.

\subsection{Effects on dust attenuation}\label{sec:dust_effects}

As mentioned above, Fig.\ \ref{fig:OPTxPHO}f bears the unexpected result that no significant change in dust attenuation is found with the addition of UV constraints. While true for the general population of galaxies, there can be important changes in $\tau_V$ for galaxies dominated by young stars. This effect gets diluted in Fig.\ \ref{fig:OPTxPHO}, and its identification requires a more careful analysis.

A useful way to evaluate the sensitivity of our attenuation estimates to UV data is to investigate the relation between the far-UV attenuation ($A_{FUV}$) and the UV spectral slope ($\beta$, assuming $F_\lambda \propto \lambda^\beta$). This relation was originally found by \cite{Meurer1999} in a study of starburst galaxies\footnote{The relation also includes the ratio between far-infrared and UV luminosities (the $IRX$ index), but this is not explored here.}, and subsequently extended to larger and more diverse samples \citep{Kong2004, Seibert2005, Buat2005, Burgarella2005}. A general conclusion from these studies is that the correlation between $A_{FUV}$ and $\beta$ found for starburst galaxies becomes much more scattered for more quiescent systems, indicating that the SFH plays an important role in the relation.

To build the relation from our data, we estimate the UV spectral slope from the $FUV-NUV$ color using $\beta_{GLX} = 2.286 \left( FUV-NUV \right) - 2.096$, as calibrated by \cite{Seibert2005}. $A_{FUV}$ is obtained from the Calzetti law, which gives $A_{FUV} = 2.53 A_V$. In order to map the effects of the SFH we subdivide the sample into three bins in \atflux, the luminosity-weighted mean $\log$ age (evaluated at 5635 \AA, as in Fig.\ \ref{fig:OPTxPHO}a).

\begin{figure}
 \centering
 \includegraphics[width=\columnwidth]{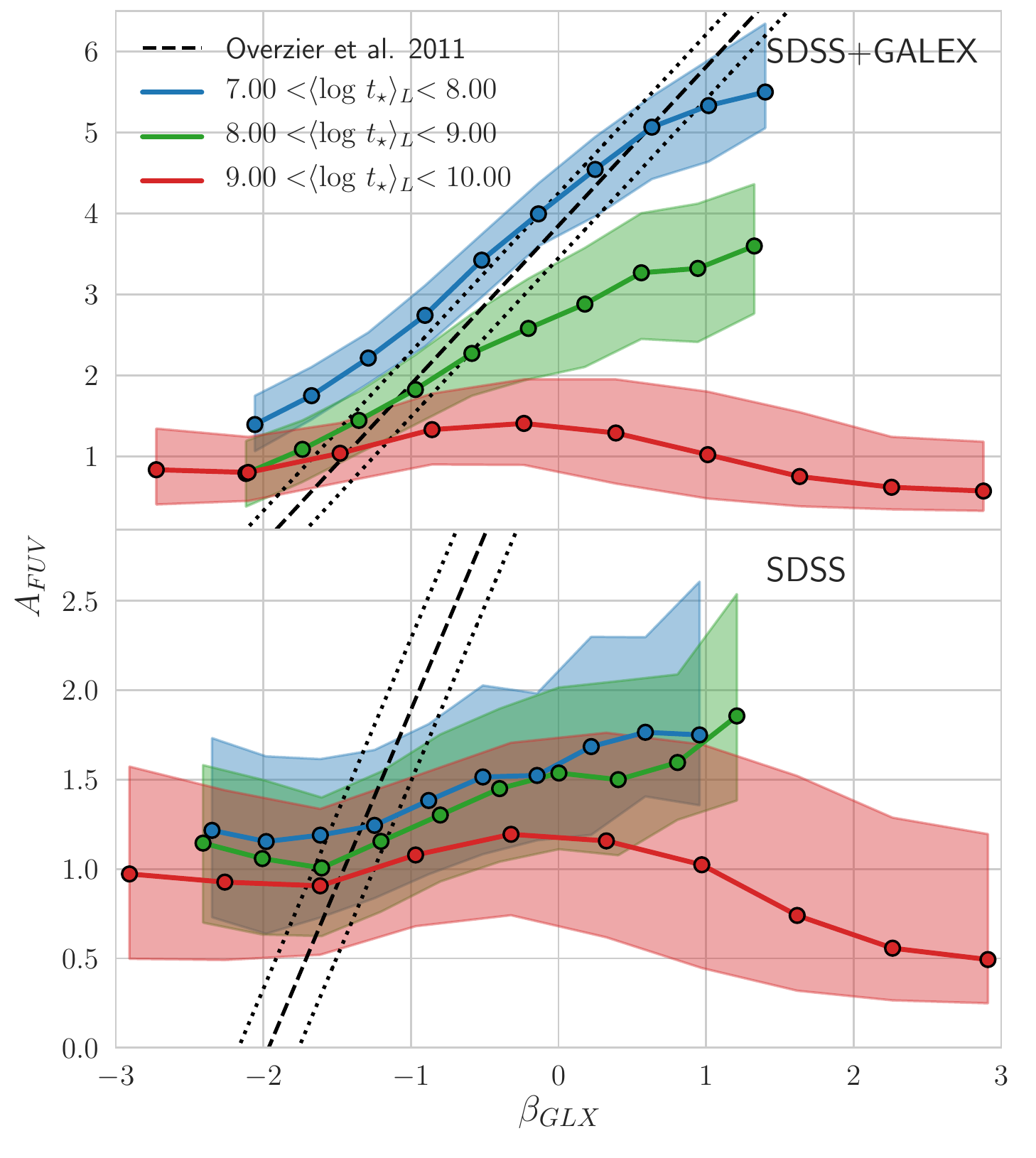}
 \caption{Relation between far-UV dust attenuation $A_{FUV}$ UV power-law slope $\beta$ for fits with (top) and without (bottom) UV data for three ranges of light-weighted mean stellar age \atflux. The relations are plotted as median curves with interquartile regions highlighted. Dotted and dashed lines represent the relation derived by \protect\cite{Overzier2011}. }
 \label{fig:A_FUV_beta}
 \end{figure} 
  
The results of this analysis are shown in Fig.\ \ref{fig:A_FUV_beta}, where $A_{FUV}$ is plotted against $\beta_{GLX}$. The top panel shows the relations for $A_{FUV}$ obtained from SDSS+GALEX fits, while the bottom panel shows estimates of $A_{FUV}$ derived from SDSS only. 
For comparison, we include the relation derived by \cite{Overzier2011}: $A_{FUV} =  3.85+1.96 \beta_{GLX} \pm 0.4$.
In fits that include UV data, the relation is well reproduced in the younger age bin (analogous to starbursts), with a smaller slope for intermediate \atflux\ systems and no correlation for galaxies dominated by old stars. Overall this confirms the previous suggestions that the SFH has a major impact on the relation between $A_{FUV}$ and $\beta$.

For SDSS-only fits, on the other hand, the relations are much weaker and more scattered, with no significant change in slope from the first to the second age bin. Note also that the values of $A_{FUV}$ estimated from SDSS+GALEX fits for star-forming galaxies are compatible with the ones presented in the literature (eg. \citealt{Meurer1999,Seibert2005,Overzier2011}), while the values obtained from SDSS-only fits are underestimated. 

Finally, another important conclusion to be drawn from Fig.\ \ref{fig:A_FUV_beta} is that changes in stellar populations are not sufficient for \textsc{starlight} to reproduce the $FUV-NUV$ colors of galaxies with significant amounts of young stars. In these cases, an increase in dust attenuation is also required.

\section{Applications}\label{sec:applications}

Having compared properties derived with and without UV constraints, this section focuses on the results obtained with our combined UV + optical synthesis analysis. The goal here is to explore some of the learning possibilities offered by this combined approach.

The broad spectral range of these fits calls for an explicit assessment of the  $\lambda$-dependence of luminosity-based descriptors of the SFH. This is the subject of section \ref{sec:SFH_x_lambda}, where the cumulative light fraction SFHs across the CMD of Fig.\ \ref{fig:SFHNUVr} are presented for a series of wavelengths from $FUV$ to $z$.
After that we address two unrelated example issues which benefit from our combined UV + optical analysis. Section \ref{sec:Lynis_x_Lineless} compares the SFHs of red-sequence ellipticals with (liny) and without (lineless) emission lines, in search for clues on why these otherwise similar galaxies differ in their emission line properties. Finally, section \ref{sec:dustlaws} explores whether our UV + optical analysis can shed light on the issue of which attenuation law best describes dust attenuation in galaxies of different types.

\subsection{Wavelength-dependent star formation histories}
\label{sec:SFH_x_lambda}

\starlight-based studies make abundant use of the luminosity-weighted mean log age $\langle \log t \rangle_L$ and SFHs expressed in terms of the light fraction population vector $\vec{x}$ (e.g \citealt{Cid2005, Abilio2006, Natalia2007, Cid2013}). Although mass-based descriptions of the SFH 
are  more directly comparable to models, light-based descriptions bear a direct and much stronger relation to observed properties which form the basis of our empirical knowledge of galaxy evolution. 

By their very definitions,  \starlight's light fractions and light-weighted mean stellar ages are $\lambda$-dependent. Section  \ref{sec:results}, for instance, presented results for $x(t)$ and $\langle \log t \rangle_L$ for $\lambda = 5635$ \AA, an arbitrary wavelength chosen for no fundamental reason other than being in a relatively clean, feature-free window. 
The purpose of this section is to take advantage of this $\lambda$-dependence, examining stellar populations in the wavelength ranges in which they are most relevant. 
Particular attention is given to the GALEX bands, as they present clear descriptions of stellar populations that are easily overlooked (or even undetected) in the optical.

\subsubsection{Spectral algebra: Converting $\vec{x}(\lambda_0)$ to $\vec{x}(\lambda)$}
\label{sec:spectralalgebra}

Consider a spectrum $L_\lambda = \sum L_{j,\lambda}$ built by superposing $j=1 \ldots N_\star$ components $L_{j,\lambda}$, and let  $x_j(\lambda_0) = L_{j,\lambda_0} / L_{\lambda_0}$ be the fractional contribution of the $j^{\rm th}$ component to the total emission at $\lambda_0$. Each $L_{j,\lambda}$ can  be written as

\begin{equation}
L_{j,\lambda} 
=  L_{j,\lambda_0} \, \left( \frac{ L_{j,\lambda}  }{ L_{j,\lambda_0} } \right) 
=  x_j(\lambda_0) \, L_{\lambda_0} \, \left( \frac{ L_{j,\lambda}  }{ L_{j,\lambda_0} } \right) 
\end{equation}

\noindent 
where $x_j(\lambda_0) \equiv L_{\lambda_0} / L_{j,\lambda_0}$ quantifies the contribution of component $j$ to the total emission at $\lambda_0$. The term in between parentheses can be expressed in terms of ratio of the intrinsic (dust-free) luminosities, $b_{j,\lambda} \equiv L^0_{j,\lambda} / L^0_{j,\lambda_0}$, and the ratio of attenuation factors at $\lambda$ and $\lambda_0$, so that

\begin{equation}
L_{j,\lambda} 
=  x_j(\lambda_0) \, L_{\lambda_0} \, b_{j,\lambda}  \, e^{-\tau_j (q_\lambda - q_{\lambda_0})}
\end{equation}

\noindent where $\tau_j$ is the V-band effective optical depth of component $j$, and $q_\lambda \equiv \tau_\lambda/\tau_V$ is given by the reddening curve. In practice $\tau_j = \tau$ if only one attenuation is allowed for, as in the case in this paper. Notice also that $b_{j,\lambda}$ are just the base spectra normalized at $\lambda_0$. 

But $L_{j,\lambda}$ can also be equated to $x_j(\lambda) L_\lambda$, where $L_\lambda$ is the total emission and $x_j(\lambda)$ is the fractional contribution of component $j$ to this total spectrum at wavelength $\lambda$. Comparing these two expressions for $L_{j,\lambda}$ leads to the sought relation between $x_j(\lambda)$ and $x_j(\lambda_0)$:

\begin{equation}
\label{eq:xlambda}
x_j(\lambda) 
= \frac{ x_j(\lambda_0) \,  b_{j,\lambda} } { \sum\limits_j x_j(\lambda_0) \,  b_{j,\lambda}},
\end{equation}

\noindent which allows one to convert the light fraction population vector from a chosen normalization wavelength $\lambda_0$ to any other $\lambda$.

\subsubsection{Wavelength-dependent SFHs in the CMD}\label{sec:xLambda}

\begin{figure*}
 \centering
 \includegraphics[width=\textwidth]{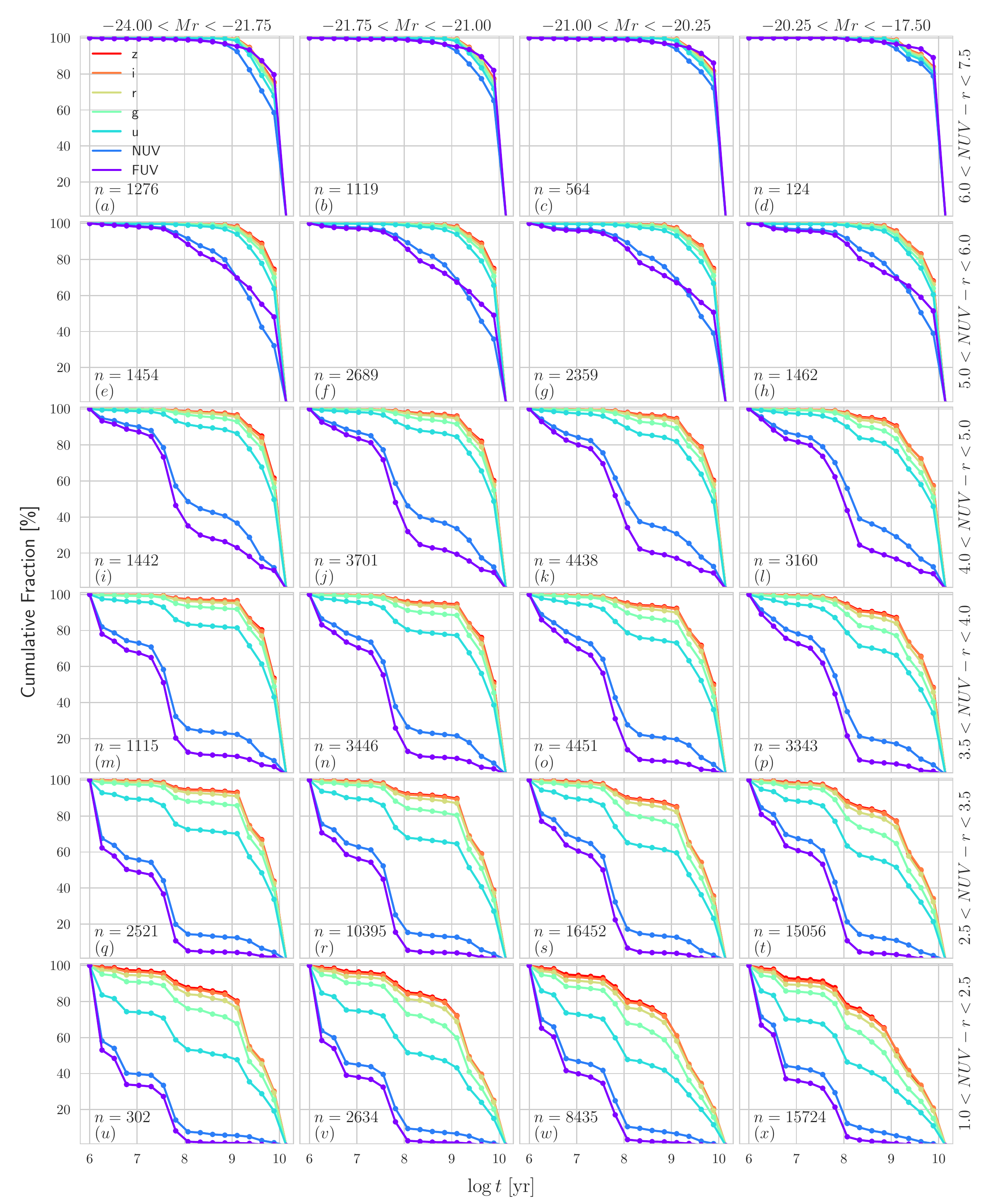}
\caption{
Average star formation histories for the same color-magnitude bins used on Fig \ref{fig:SFHNUVr}, expressed as cumulative light fractions $x(>t,\lambda)$ in wavelengths corresponding to SDSS and GALEX bands. 
All curves correspond to results obtained in the combined GALEX+SDSS \starlight\ fits. 
 }
 \label{fig:xOfLambda_x_t}
 \end{figure*}

With equation \ref{eq:xlambda}, we calculated light fractions for all GALEX and SDSS bands. Fig.\ \ref{fig:xOfLambda_x_t} shows the $x(>t,\lambda)$ cumulative light fractions as a function of age, as derived from our combined GALEX+SDSS fits. These $\lambda$-dependent descriptions of the SFH are broken into the same bins in the UV-optical CMD as in Fig.\ \ref{fig:SFHNUVr}. Colors code for the reference wavelength, as labeled in the top-left panel.

The $x(>t,\lambda)$ curves for the longer wavelengths like the r, i, and z bands 
rise faster for more massive galaxies (left panels) than for those of lower mass (right), revealing the usual downsizing pattern. Because of the small influence  of young stars at these wavelengths these curves essentially reflect the mass growth curves previously shown in Fig.\ \ref{fig:SFHNUVr}. At UV wavelengths, on the other hand, the light is always dominated by $< 10^8$ yr populations, and the curves are similar for galaxies of different masses. Relevant changes in the $NUV$ and $FUV$ curves occur only in the vertical direction, with the proportion of $10^{6-7}$ to $10^{7-8}$ populations decreasing as $NUV - r$ becomes redder.

Panels (i) to (l) show galaxies in the green valley, mostly populated by an intermediate population of recently quenched galaxies moving from the blue cloud to the red sequence. Accordingly, all $x(>t,\lambda)$ curves shift towards larges ages. Unlike in the blue cloud, $> 10^8$ yr populations now contribute significantly even at UV wavelengths, particularly for the more massive galaxies.

Red sequence galaxies are shown in the top two rows of Fig. \ref{fig:xOfLambda_x_t}. For galaxies in panels (e) to (h),  intermediate age populations are still relevant, while galaxies in the upper red-sequence (top row, panels a to d) are completely dominated by old stellar populations, even in the UV.

The evolutionary synthesis models used in this work become redder in $FUV-NUV$ up to stellar ages of 1 Gyr. After that the model spectra become bluer in the UV, and after about 4 Gyr the stellar populations emit more $FUV$ per $M_\odot$ than immediately younger populations. This results in a rise on the contribution of the oldest populations to $FUV$ light, surpassing their contribution to the $NUV$. This tendency is connected to the UV upturn phenomenon, since at constant $NUV-r$ the contribution of old populations to $FUV$ light correlates with $FUV-NUV$ color.

\subsection{Liny and lineless retired galaxies}
\label{sec:Lynis_x_Lineless}

\begin{figure*}
 \centering
 \includegraphics[width=\textwidth]{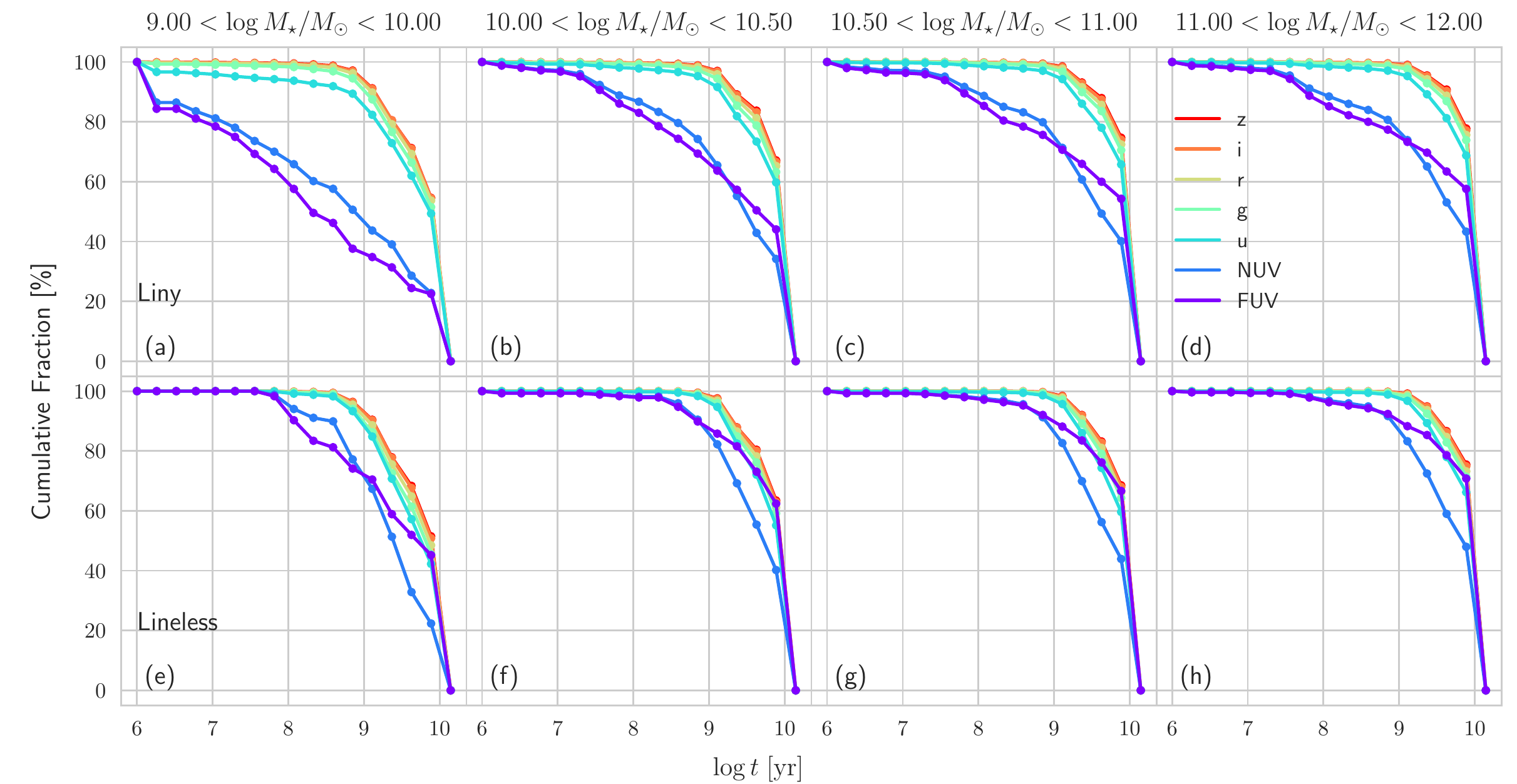}
 \caption{Wavelength dependent SFHs for liny (top row) and lineless (bottom row) retired galaxies divided into four mass ranges. Color codes are the same as in Fig \ref{fig:xOfLambda_x_t}.
 }
 \label{fig:elr_llr}
 \end{figure*}

Up until now, we have used the UV-optical color-magnitude diagram as
our guide to subdivide our sample into intelligible sub-classes. This diagram provides 
information on galaxy evolution and the
bimodality of the local galaxy population. However, galaxies are complex systems, and no single diagram condenses all the information needed
to describe them. Another popular way to divide the galaxy population is based on emission line diagrams. This type of classification
separates galaxies with different sources of ionizing photons, an
aspect that hasn't yet been touched in this work.

A popular emission line based classification is the one based on the $\Nii/\Ha$ flux ratio and the \Ha\ equivalent width ($W_{\rm H\alpha}$), the so called WHAN diagram, introduced by \cite{Cid2011}. The main advantage of the WHAN diagram over those based solely on line ratios (e.g. \citealt*{BPT}) is that it can identify retired galaxies \citep{Grazyna2008,Grazyna2015}. 
This diagram allows the identification of retired galaxies, characterized by having $W_{\rm H\alpha} < 3$\,\AA. This class of galaxies can be misclassified as AGN in other diagrams, even though their ionization field is dominated by HOt Low Mass Evolved Stars (HOLMES) typical of old stellar populations.

Retired galaxies can be further subdivided into liny ($0.5 < W_{\rm H\alpha} < 3$\AA) and lineless ($W_{\rm H\alpha} < 0.5$\AA) systems, according to the presence or absence of emission lines. The fact that some retired galaxies lack emission lines poses an interesting astrophysical problem, since the HOLMES in both subclasses produce ionizing photons enough to power line emission.
This problem is studied in detail by \citet[][hereafter H18]{Herpich2018}. The aim of this section is to complement their analysis using our UV--optical spectral synthesis.

Through a pair-matching analysis, H18 were able to identify very small but consistent differences between the two sub-classes of retired galaxies. Compared to lineless systems, liny ones tend to be brighter in the GALEX $NUV$ and WISE $W3$ bands, and have slightly smaller 4000\,\AA\ breaks. These results are indicative of differences in dust content and SFHs.
These differences in SFH, however, are hard to measure directly in the optical.

We have used our combined GALEX+SDSS spectral analysis to address this issue in the hope that a wider wavelength range leads to a clearer separation between the SFHs of liny and lineless retired galaxies. To this end, we have culled a sample of retired galaxies out of our low-$z$ sample. In addition to the $W_{\rm H\alpha}$ selection we also require the galaxies to be classified as ellipticals by Galaxy Zoo. 
Out of the 16206 retired galaxies in our sample, 12\% are lineless and the remaining are liny.

Fig.\ \ref{fig:elr_llr} shows mean cumulative light fraction curves for lineless (top panels) and liny (bottom panels) galaxies, split into four bins of stellar mass. As in Fig. \ref{fig:xOfLambda_x_t}, each panel shows the $x(>t)$ curves for seven wavelengths corresponding to the GALEX and SDSS filters. Variations in the curves from left to right panels reflect the well known downsizing pattern, with less massive galaxies having SFHs that are more extended in lookback time. 
The SFHs for the UV bands clearly show that liny galaxies experienced a more extended period of star-formation, a trend that is consistent in all mass ranges. The same signal is present in the $x(>t)$ curves for optical wavelengths, but at a much reduced amplitude.

Though populations of a few Myr may be present in low mass liny galaxies, the main difference is an excess of intermediate age populations ($\sim 0.1$--1 Gyr). 
This directly shows that the presence or absence of emission lines in retired galaxies is connected to slightly different SFHs.
H18 interprets this as due to an external reservoir of cold gas that is slowly accreted by liny galaxies, feeding a low level of star-formation over an extended period of time.
In contrast, lineless systems exhausted their gas supply at an early epoch, leaving little or no gas to react to the ionizing radiation field produced by their HOLMES.

Another difference between these galaxy classes is the larger dust content in liny galaxies. The optical-only analysis on H18 is already capable of identifying higher $\tau_V$ values for liny galaxies at all mass ranges, a result that is confirmed here  for both optical and UV+optical fits. Consistently with the H18 findings, we identify an average difference of $\Delta \tau_V = 0.15$ between liny and lineless galaxies.

\subsection{Comparing results for different attenuation laws}\label{sec:dustlaws}

Previous experiments with \starlight\ found that spectral fits in the optical cannot distinguish among different choices of dust attenuation law. \cite{Natalia2007}, for instance, show that equally good fits are obtained with $q_\lambda=A_\lambda/A_V$ functions representative of the Milky-Way, Magellanic Clouds, or starburst galaxies. This degeneracy is hardly surprising, given that the main differences between these alternative laws lies not in the optical, but in the UV.

One of the advantages of incorporating UV data in the \starlight\ analysis is that it allows us to revisit this issue and investigate which dust attenuation law best fits the data. In this section we compare results obtained with the CAL and CCM laws, whose most notable difference is presence or absence of the UV bump at $\lambda \sim 2175$ \AA.
Given the overlap with the $NUV$ filter, this feature is bound to affect our analysis. $FUV$ fluxes are also affected by the choice of $q_\lambda$, given that the CCM law have a steeper far-UV slope, even though slopes in the optical are similar.
Though the CAL and CCM laws do not span the wealth of dust attenuation effects studied in the literature (see \citealt{CIGALE,Seon2016,Salim2018,Desika2018}), they serve as useful limits to investigate the effect of the bump. Moreover, they are both commonly used in \starlight-based studies.

Qualitatively one expects fits using a CAL law to produce larger residuals in the UV than fits with a CCM law for galaxies with relevant UV bumps, and vice versa. 
It is therefore interesting to compare the UV residuals obtained with these two laws. To evaluate the quality of the fits in the UV let us use 

\begin{equation}
 \chi_{\rm UV} = \sqrt{\frac{1}{2} \chi^2_{\rm PHO}},
\end{equation}
 
\noindent which gives an average of the $NUV$ and $FUV$ reduced residuals.

For this analysis we will restrict the sample to galaxies with $NUV-r<5$ that are classified as spirals by Galaxy Zoo. On the other hand we do allow systems beyond $z=0.1$ in order to trace the effect of the $NUV$ band moving out of the UV bump region as redshift increases. These cuts yield a sample of 81214 galaxies. 

\begin{figure}
 \centering
 \includegraphics[width=\columnwidth]{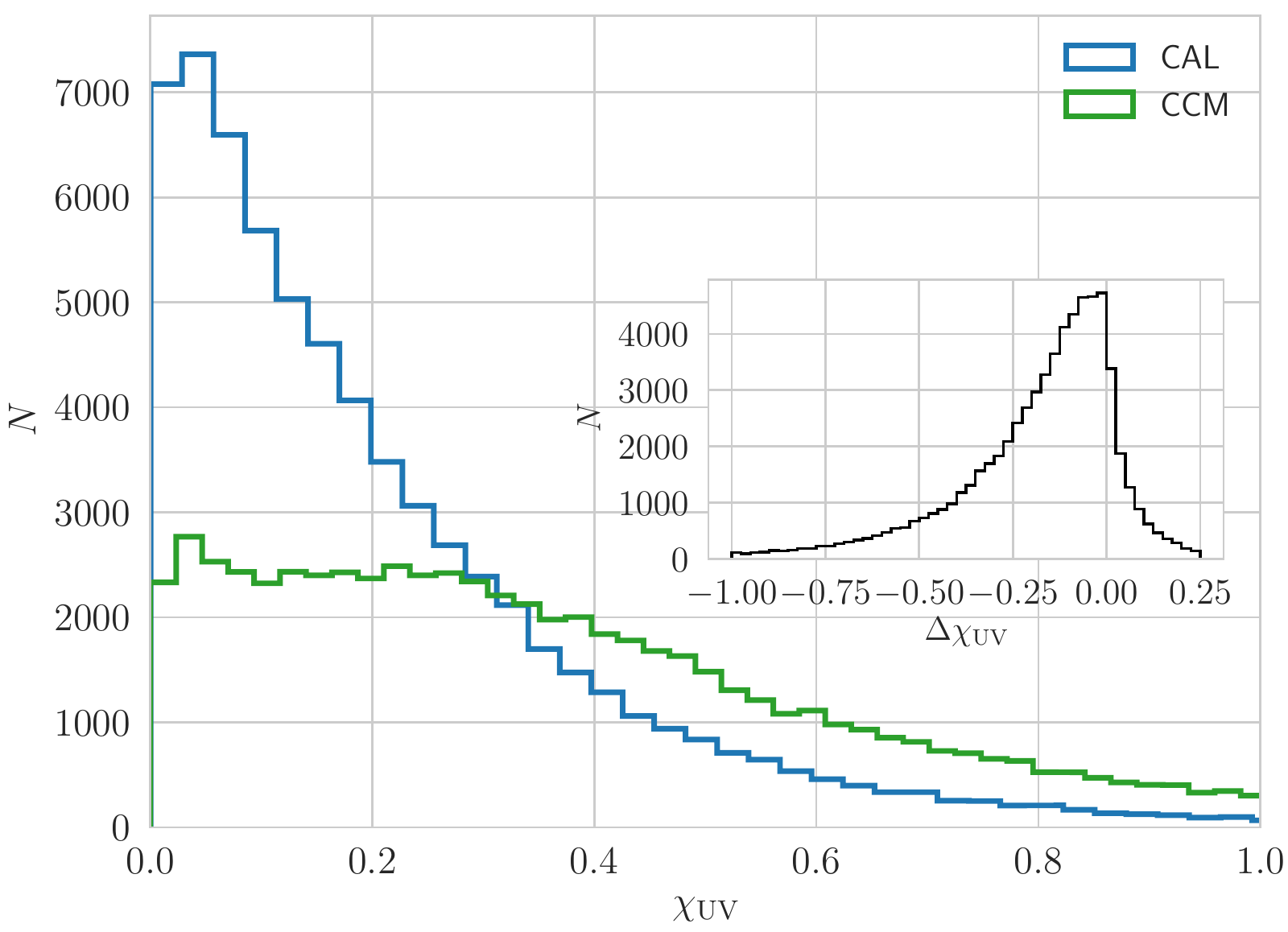}
 \caption{Histograms of $\chi_{\rm UV}$ for the CAL attenuation  (blue) and CCM extinction (green) laws.
The inset shows the distribution of $\Delta\chi_{\rm UV} = \chi^{\rm CAL}_{\rm UV} - \chi^{\rm CCM}_{\rm UV}$. Galaxies in these histograms are all Galaxy Zoo spirals with $NUV - r < 5$.  
 }
 \label{fig:extlaws_hist}
 \end{figure}

Fig. \ref{fig:extlaws_hist} shows histograms of $\chi_{UV}$ for the CAL and CCM laws. The mean and rms values of $\chi_{\rm UV}$ are $0.22\pm0.26$ for the CAL law and $0.41\pm0.36$ for CCM.
The inset shows the histogram of $\Delta\chi_{\rm UV} = \chi^{\rm CAL}_{\rm UV} - \chi^{\rm CCM}_{\rm UV}$. Over our whole sample $\Delta\chi_{\rm UV}$ averages to $-0.18$, showing that the CAL law is generally preferred over CCM.

\begin{figure}
 \centering
 \includegraphics[width=\columnwidth]{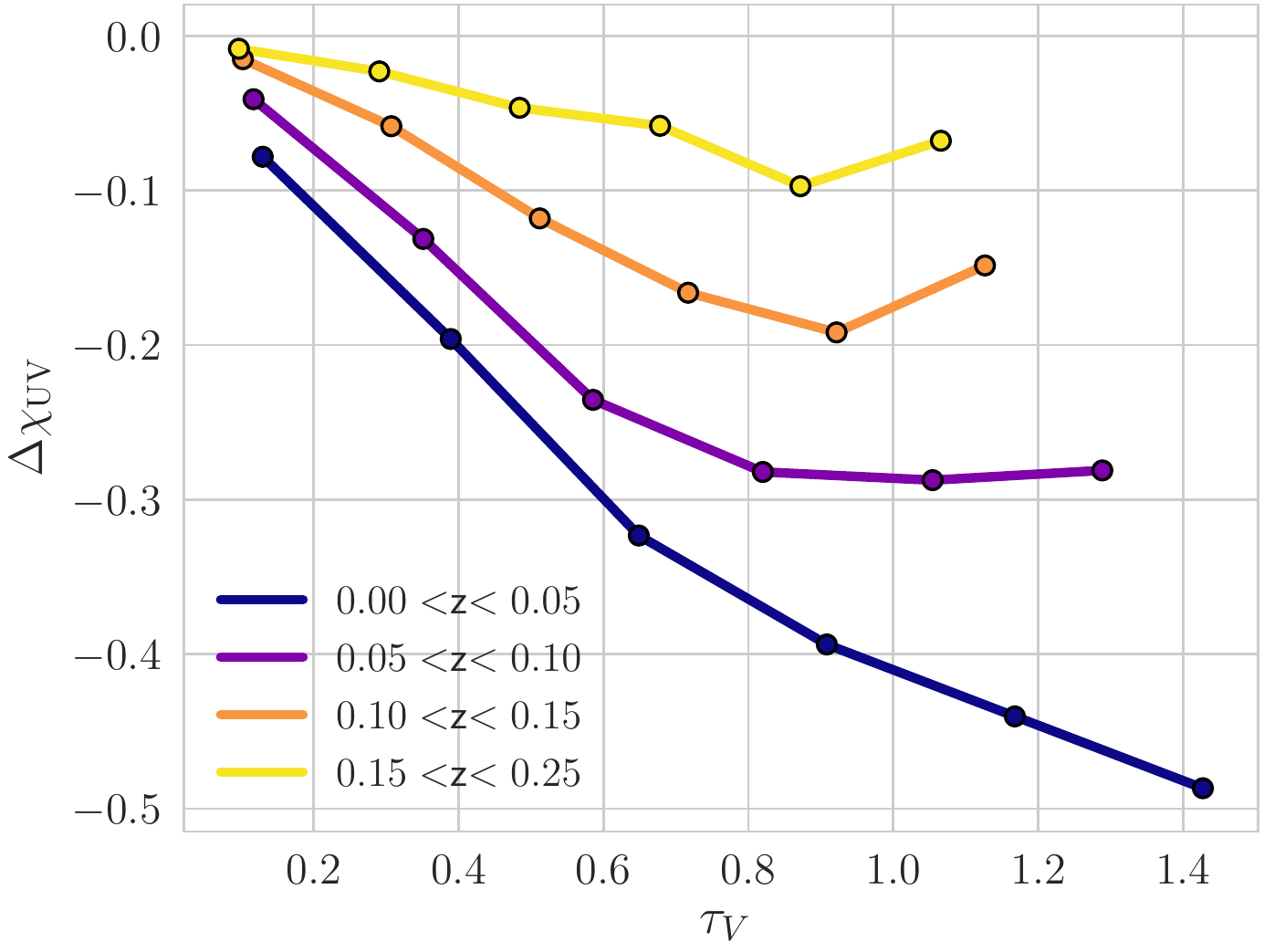}
 \caption{Median $\Delta\chi_{\rm UV}$ plotted against dust optical depth $\tau_V$ in different bins of redshift. The center of the bins used to calculate the medians are indicated as points.}
 \label{fig:bump}
 \end{figure} 
 
We find that the main parameters controlling  how worse the fits get with the CCM law are $\tau_V$ and redshift. 
Trends with mass and star-formation rate were also investigated, but we found no visible dependence of $\Delta\chi_{\rm UV}$ with these parameters at fixed $\tau_V$ and redshift.
In Fig. \ref{fig:bump}, we plot median curves of $\Delta\chi_{\rm UV}$ as a function of $\tau_V$ for four different ranges in $z$. The $\tau_V$ values used in this plot are those derived from purely optical fits with a CAL attenuation law, which are independent from the UV. The plot shows that, in comparison with CCM, the CAL law yields progressively better fits of the GALEX photometry as $\tau_V$ increases.
The systematic behavior shown in Fig. \ref{fig:bump} is expected to a certain degree, since the bump level is amplified as $\tau_V$ increases. Moreover, the quasi-universal relation between $\tau_V$ and the slope of the attenuation curve at any wavelength found by \cite{Chevallard2013} (and confirmed by \citealt{Salim2018}) also takes the results in this direction by favoring a steeper slope (MW-like) for low $\tau_V$. 
The effects of redshift are also evident in this plot. At fixed $\tau_V$, the $\Delta\chi_{\rm UV}$ is largest (in modulus) at low redshift (blue curves in Fig. \ref{fig:bump}). As redshift increases the CAL and CCM laws yield increasingly similar $\chi_{\rm UV}$. We interpret this as due to the gradual shifting of the UV bump away of the NUV band as redshift increases. To first order this makes the two laws similarly bump-less, although differences in far-UV slope between the two $q_\lambda$ functions persist.

Overall, the results in Fig. \ref{fig:bump} indicate that the attenuation curves of the general population of spiral galaxies either lack the features that distinguish the MW extinction curve or exhibit them at a less significant level.
This is compatible with other studies that generally find a small level of the bump both locally \citep{CSB, Wild2011, Salim2018}, and at higher redshift \citep{Kriek2013,Reddy2015}. For instance, \cite{Buat2011} find an average bump amplitude of 35\% the MW value for galaxies at $z>1$.  
Similarly to our results, \cite{Battisti2016} find that the population of local star-forming galaxies can be described by a CAL-like law, but small levels of the bump cannot be discarded.
This is not indicative, however, that the dust grain population in the MW is somewhat unusual, as recent models show that a bumpless attenuation law can arise even when the underlying extinction curve is MW-like \citep{Desika2018}.

\begin{figure*}
 \centering
 \includegraphics[width=\textwidth]{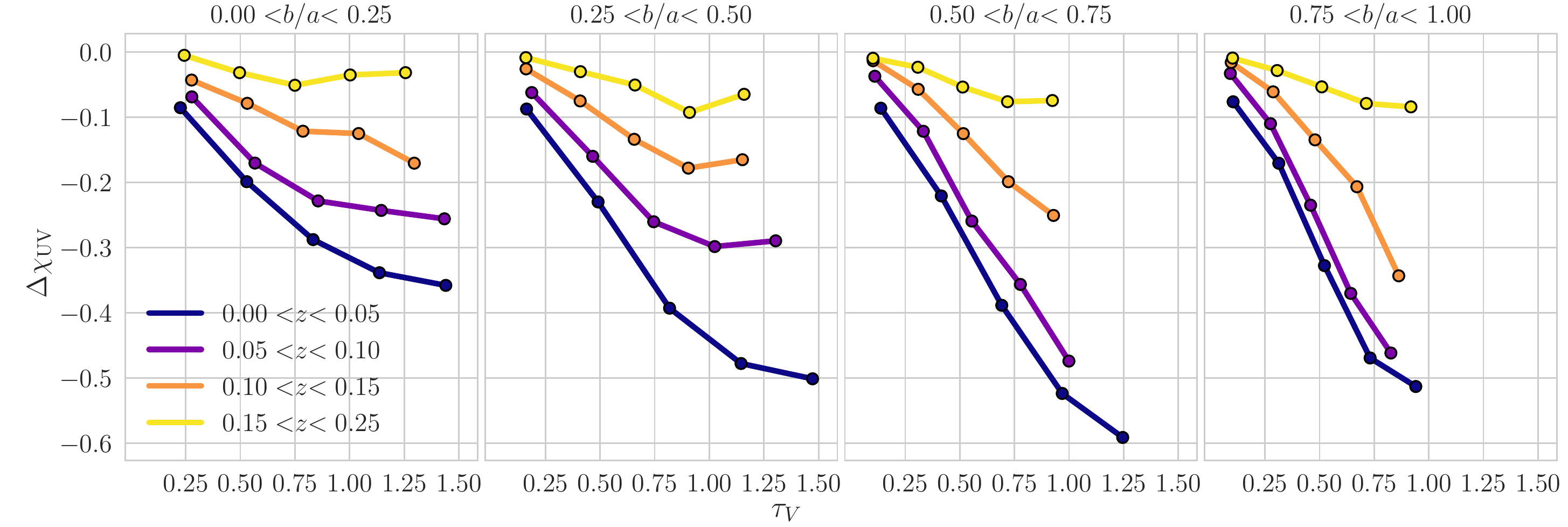}
 \caption{Median $\Delta\chi_{\rm UV}$ plotted against dust optical depth $\tau_V$ in different bins of redshift in four ranges of inclination $b/a$. Edge-on galaxies are shown in the left and face-on galaxies in the right.}
 \label{fig:bump_ba}
 \end{figure*} 

Identifying a population of galaxies with relevant UV-bumps in our data-set is not straightforward, since the relation of $\Delta\chi_{\rm UV}$ with attenuation is so strong that it contaminates other correlations. For instance, a key parameter in this analysis is galaxy inclination, but, since $b/a$ correlates strongly with attenuation, plotting it against $\Delta\chi_{\rm UV}$ would only show the reverberation of the correlation of both variables with $\tau_V$. 
An attempt to isolate the effects of galaxy inclination in the shape of the attenuation curve is shown in Fig. \ref{fig:bump_ba}, where we repeat Fig. \ref{fig:bump} splitting the sample in four ranges of $b/a$. Edge-on galaxies are shown on the left and face-on ones on the right. As in Fig. \ref{fig:bump}, we are unable to identify a significant population of galaxies that is best fitted with the CCM law. That aside, a trend is clear in Fig. \ref{fig:bump_ba}: fits with the CCM law get worse with decreasing inclination (increasing $b/a$). 

This goes in the direction of the results of \cite{CSB}, although in that work it is found that a MW-like extinction law is required to fit the UV spectral slope of highly inclined galaxies, while we find only that these are the galaxies where MW law works best. Even in these cases our results are better for the CAL law. 
The relation of $\Delta\chi_{\rm UV}$ with inclination is consistent with the idea that the shape of the MW extinction curve in the UV is associated with diffuse dust \citep{Wild2011}, while the CAL law is best-suited for birth clouds. As galaxy inclination increases, the amount of dust from birth clouds in our line of sight stays constant, while the column of diffuse dust increases. In this interpretation, highly inclined galaxies should show attenuation curves with shapes closer to the MW extinction law, as seen in Fig. \ref{fig:bump_ba}. 
However, this result is not universal. For instance, the radiative transfer models of \cite{Pierini2004} show a decrease in the bump for edge-on galaxies.

Our analysis is sufficient to showcase the potential of combined SDSS+GALEX \starlight\ fits in distinguishing between dust attenuation laws. 
A more thorough analysis, exploring laws with different slopes and bump strengths (e.g., \citealt{CIGALE}), as well as fits with a population-dependent dust attenuation (\citealt{Charlot2000}) is deferred to a future study.

\section{Summary}\label{sec:conclusions}

We have developed a method to estimate GALEX UV magnitudes in apertures consistent with SDSS spectra. Tests indicate that this method produces very little bias, but considerable scatter, making it suitable for statistical analysis of large datasets, though not for individual sources.
We also improved upon the version of \starlight\ used by \cite{Rafa2016}, introducing a new method for combining spectroscopic and photometric figures of merit. Other improvements on the code remain under the hood as we work towards a public distribution of this new version of \starlight.

We use the new code to simultaneously analyze SDSS spectra and GALEX photometry, retrieving stellar population properties of 231643 galaxies. Our overall results agree with previous work based on CALIFA+GALEX data, with few exceptions introduced by differences in stellar population models and spectral coverage. 
As seen in Fig.\ \ref{fig:example_fits}, with the addition of UV constraints \starlight\ has to bring UV fluxes down without producing a redder optical spectrum. To achieve this, the code attributes larger light fractions to stellar populations with $10^7$ to $10^8$ yr, while cutting from the contribution of very young and very old stars to maintain optical colors.
This behavior produces slightly older mean stellar ages when weighted by light, and slightly younger when weighted in mass, as seen in Fig.\ \ref{fig:OPTxPHO}. We also find an increase in dust attenuation for galaxies dominated by young stars (Fig.\ \ref{fig:A_FUV_beta}).

The panchromatic nature of our synthesis results was explored in section \ref{sec:SFH_x_lambda} by calculating wavelength-dependent star formation histories. As expected, UV light fractions are good tracers of recent star formation in the blue cloud, while optical bands are mostly tracers of old stellar populations. We also identify that UV light fractions are able to distinguish the stellar populations of liny and lineless retired galaxies, classes whose small differences in SFH are barely noticeable in the optical.
In section \ref{sec:dustlaws}, we showcase our ability to distinguish between attenuation laws of different shapes, while keeping a non-parametric analysis of stellar populations. Our results for the law of starburst galaxies of \cite{Calzetti2000} are systematically  better than the ones for the MW law of \cite{CCM}.
A complete study on this topic, however, would have to include laws of different slopes and bump strengths.

The methodology of this paper provides galaxy properties derived from non-parametric spectral synthesis applied to optical spectra and UV photometry, expanding on previous \starlight-based studies.
The full potential of the data-set provided by our analysis is beyond what is explored in this paper, where we only presented short case-studies. Future applications may include a study of the UV upturn in elliptical galaxies and a deeper analysis of the shape of dust attenuation curves, as well as the effects of binary stars and differential dust attenuation to the UV.

\section*{Acknowledgements}

The authors thank the anonymous referee for the valuable comments about the paper and especially for inspiring us to write section \ref{sec:dust_effects}.
AW would like to thank Andrew Battisti for the valuable discussions about GALEX+SDSS aperture matching schemes. 
AW, RCF, NVA and FRH acknowledge the support from the CAPES CSF--PVE project 88881.068116/2014-01. 
This study was financed in part by the Coordena\c{c}\~ao de Aperfei\c{c}oamento de Pessoal de N\'{\i}vel Superior -- Brasil (CAPES) -- Finance Code 001. NVA acknowledges support of the Royal Society and the Newton Fund via the award of a Royal Society--Newton Advanced Fellowship (grant NAF\textbackslash{}R1\textbackslash{}180403), and of FAPESC and CNPq. RGD thanks support from the Spanish Ministerio de Econom\'ia y Competitividad through the project AyA2016-77846-P.
GB acknowledges financial support through PAPIIT project IG100115 from DGAPA-UNAM.
Funding for the SDSS and SDSS-II has been provided by the Alfred P. Sloan Foundation, the Participating Institutions, the National Science Foundation, the U.S. Department of Energy, the National Aeronautics and Space Administration, the Japanese Monbukagakusho, the Max Planck Society, and the Higher Education Funding Council for England. The SDSS Web Site is \url{http://www.sdss.org/}.
The SDSS is managed by the Astrophysical Research Consortium for the Participating Institutions. The Participating Institutions are the American Museum of Natural History, Astrophysical Institute Potsdam, University of Basel, University of Cambridge, Case Western Reserve University, University of Chicago, Drexel University, Fermilab, the Institute for Advanced Study, the Japan Participation Group, Johns Hopkins University, the Joint Institute for Nuclear Astrophysics, the Kavli Institute for Particle Astrophysics and Cosmology, the Korean Scientist Group, the Chinese Academy of Sciences (LAMOST), Los Alamos National Laboratory, the Max-Planck-Institute for Astronomy (MPIA), the Max-Planck-Institute for Astrophysics (MPA), New Mexico State University, Ohio State University, University of Pittsburgh, University of Portsmouth, Princeton University, the United States Naval Observatory, and the University of Washington. This project made use of GALEX data and the Barbara A. Mikulski Archive for Space Telescopes.

\bibliographystyle{mnras}
\bibliography{references} 







\label{lastpage}

\appendix

\end{document}